\documentclass[10pt,english,aps,prd,superscriptaddress,notitlepage,nofootinbib,preprintnumbers,floatfix]{revtex4-1}
\usepackage[latin9]{inputenc}
\usepackage{amsmath}
\usepackage{graphicx}

\makeatletter

\providecommand{\tabularnewline}{\\}

\usepackage{babel}

\usepackage{babel}

\usepackage{babel}

\usepackage[caption=false]{subfig}
\makeatother

\usepackage{babel}
\begin{document}
\preprint{YITP-19-52, IPMU19-0088}
\title{Baryon Physics and Tight Coupling Approximation in Boltzmann Codes}
\author{Masroor C.~Pookkillath}
\email{masroor.cp@yukawa.kyoto-u.ac.jp}

\affiliation{Center for Gravitational Physics, Yukawa Institute for Theoretical
Physics, Kyoto University, 606-8502, Kyoto, Japan}
\author{Antonio De~Felice}
\email{antonio.defelice@yukawa.kyoto-u.ac.jp}

\affiliation{Center for Gravitational Physics, Yukawa Institute for Theoretical
Physics, Kyoto University, 606-8502, Kyoto, Japan}
\author{Shinji Mukohyama}
\email{shinji.mukohyama@yukawa.kyoto-u.ac.jp}

\affiliation{Center for Gravitational Physics, Yukawa Institute for Theoretical
Physics, Kyoto University, 606-8502, Kyoto, Japan}
\affiliation{Kavli Institute for the Physics and Mathematics of the Universe (WPI),
The University of Tokyo, Institutes for Advanced Study, The University
of Tokyo, Kashiwa, Chiba 277-8583, Japan}
\begin{abstract}
 We provide two derivations of the baryonic equations that
  can be straightforwardly implemented in existing Einstein--Boltzmann
  solvers. One of the derivations begins with an action principle,
  while the other exploits the conservation of the stress-energy
  tensor. While our result is manifestly covariant and satisfies the
  Bianchi identities, we point out that this is not the case for the
  implementation of the seminal work by { Ma and
    Bertschinger and in the existing} Boltzmann codes. We also study
  the tight coupling approximation up to the second order without
  choosing any gauge using the covariant full baryon equations. We
  implement the improved baryon equations in a Boltzmann code and
  investigate the change in the estimate of cosmological parameters by
  performing an MCMC analysis. With the covariantly correct baryon
  equations of motion, we find $1\%$ deviation for the best fit values
  of the cosmological parameters that should be taken into
  account. While in this paper, we study the $\Lambda$CDM model only,
  our baryon equations can be easily implemented in other models and
  various modified gravity theories. 
\end{abstract}
\maketitle
\section{Introduction}

Studies in modern cosmology heavily rely on cosmological linear
perturbation theory \cite{Kodama:1985bj,Dodelson:2003ft}. To
understand the evolution of these linear perturbations, Einstein's
equations are solved numerically at the linear order in perturbations
around a homogeneous and isotropic background. These numerical solvers
are often referred to as Boltzmann solvers or Boltzmann codes. There
are several open-source linear Boltzmann solvers available, namely
{the Cosmic Linear Anisotropy Solving System (CLASS)}
\citep{Blas:2011rf}, {Code for Anisotropies in the Microwave
  Background (CAMB)} \citep{Lewis:1999bs}, {CMBEASY}
\citep{Doran:2003sy}, {CMBFAST} \citep{Seljak:1996is}, etc.  Among
them, CAMB and CLASS are maintained frequently. These codes provide us
with a platform to test any theory against observations.
appropriate

To understand the nature of the dark sector of our universe, there are
several future experiments planned, such as {EUCLID}
\citep{Laureijs:2011gra}, {the Dark Energy
  Spectroscopic Instrument (DESI)}
\citep{Aghamousa:2016zmz}, and {Large Synoptic Survey Telescope
  (LSST)}
\citep{Zhan:2017uwu}. All these
experiments focus on higher precision than the current dataset like
Planck, for the estimate of cosmological parameters. In order to take
full advantage of these experiments, the Boltzmann codes need to be
precise enough, and any inconsistencies present in the implementation
of the codes need to be fixed. Otherwise, theoretical predictions
cannot be matched with the observations.

The first calculation of cosmological perturbation theory was
performed by Lifshitz \citep{Lifshitz:1945du}. Later, Bardeen
\citep{Bardeen:1980kt} and Kodama and Sasaki \citep{Kodama:1985bj}
fixed the gauge issues in the scalar sector. CMBFAST
\citep{Seljak:1996is} makes use of the line-of-sight integration
method to compute anisotropies, and their code was made publicly
available. This could reduce the time for computation up to two
times. There were two other Boltzmann codes available before, one
developed by Sugiyama \citep{Sugiyama1992,Sugiyama:1994ed}, based on
gauge invariant formalism, and the other developed by White, based on
the synchronous gauge \citep{White:1995qm,Hu:1997hp,Hu:1997mn}.
CMBFAST is also based on the synchronous gauge. CMBEASY and CAMB are
basically formulated based on CMBFAST. The Boltzmann equations in
CMBFAST are taken from {COSMIC} 
\citep{Bertschinger:1995er}, which is based on the seminal work by Ma
and Bertschinger \citep{Ma:1995ey}.  {CLASS} 
is also based on it, implemented in Newtonian and synchronous gauges.

However, in \citep{Ma:1995ey}, there appeared to be at least three
possible problems in the evolution equation for the baryon fluid.
\begin{itemize}
\item First, there is a gauge incompatibility, that is the equations
  break general covariance. In particular, the equations of motion in
  the Newtonian gauge and those in the synchronous gauge are not
  related to each other by a gauge transformation. This results in
  different physical outcomes for different gauge choices. The
  consequence is that it is not clear which gauge one should choose
  from the beginning to study baryons. This should not happen in a
  covariant theory, as a gauge choice merely represents a choice of
  coordinates and does not affect physical results.
\item Second, it breaks the Bianchi identity. This aspect also leads to
inconsistencies. For example, breaking the Bianchi identity implies that
solving all components of the Einstein equations would lead in general
to a solution that is not consistent with the conservation equation
for the matter fields. 
\item Third, in the limit of no interaction between the baryon fluid and
the photon gas, we have an equation of motion for the baryons in which
the squared sound speed $c_{s}^{2}$ is present. It is difficult to
understand the nature of this term with $c_{s}^{2}$, as no known
covariant action for matter would lead to such a term in the dynamical
equation of motion. 
\end{itemize}
As mentioned above, these equations are taken for code implementation
in almost all existing Boltzmann solvers. We feel that in this era
of precision cosmology, the Boltzmann solvers should have these issues
fixed. These issues cause artificial deviations from general relativity.
These problems appeared also in \citep{Hu:1997hp}. We find that there
are terms missing at the order of $c_{s}^{2}$ in the baryon evolution
equation. In this work, we derive the correct equations of motion
for baryons from an action. The resulting equations are devoid of
the above mentioned issues. We shall see that our new terms do not
modify strongly the final results (and this is a bit reassuring);
nonetheless, we believe our corrections are to be made for Boltzmann
solvers, especially at early times, when the sound speed of the baryon
fluid (though small) cannot be neglected completely.

In fact, in the regime before recombination, photons and baryons coupled
to form a stiff baryon-photon fluid. Since, in this case, the equations
for this photon-baryon fluid are numerically stiff, in order to study
this regime, both CLASS and CAMB make use of the so-called tight coupling
approximations. These were first developed by Peebles and Yu \citep{Peebles:1970ag}.
In \citep{Ma:1995ey}, the authors developed this approximation
up to the first order, and they further assumed $\tau_{c}\propto a^{2}$
(where $\tau_{c}^{-1}\equiv an_{e}\sigma_{T}$, $a$ is the scale
factor, $n_{e}$ is the electron density, and $\sigma_{T}$ is the Thomson
cross-section) and $c_{s}^{2}\propto a^{-1}$ (where $c_{s}^{2}$
is the speed of propagation for the baryon fluid). The full first
order tight coupling approximation was implemented in \citep{Lewis:1999bs}
with an additional approximation, $\tau_{c}\propto a^{2}.$ For CMBEASY,
tight coupling approximation is calculated for the Newtonian gauge
in which they considered terms beyond the first order \citep{Doran:2003sy}.
CLASS developed both first order and second order tight coupling approximation
without assuming any further assumptions like in COSMIC and CAMB.
Several other authors have worked on the tight coupling approximation.
For example, the tight coupling approximation up to the second order
was implemented for the calculation in the synchronous gauge in \citep{CyrRacine:2010bk}.
An extension of the tight coupling approximation to second order in
the perturbation was developed in \citep{Pitrou:2010ai}. The approximation
depends on the baryon equations. Hence, also the approximation methods
used to solve the baryon-photon system need to be revised.

In order to fix the covariance issues, we find it useful to understand
the baryon physics, starting from a newly written Lagrangian, which
up to a field redefinition is equivalent to the Schutz--Sorkin Lagrangian
\citep{Schutz:1977df}, which was studied also in \citep{Brown:1992kc,DeFelice:2009bx}
in the context of perfect fluids, with general equations of state.
We then consider the baryon fluid as an ideal gas. This is in fact
a gas model for non-relativistic particles with a non-zero speed of
propagation, i.e.,\ $c_{s}^{2}\neq0$. This system then allows us
to find covariant equations of motion for the perturbation variables
(we also give an alternative derivation of the same set of equations
based on the conservation of the stress-energy tensor). We find that
there is an extra term of the order of $c_{s}^{2}$ in both the evolution
of energy density and velocity perturbations for the baryon fluid.
This fix solves all three problems mentioned above. We then use
these new baryon equations of motion in order to derive the tight
coupling approximation equations up to the second order. We implement
these corrections for the baryon evolution, in the CLASS code. Finally,
we make a parameter estimation for the $\Lambda$CDM model with these
new corrections, using the Monte Carlo sampler Monte Python \citep{Audren:2012wb,Brinckmann:2018cvx}.
We find that the new equations, solved by CLASS, give some deviations
from the previous results, but for the parameter estimation of the
$\Lambda\text{CDM}$ model, such deviations are inside the statistical
error bars (on the other hand, we do not know whether the deviation
remains within the statistical error bars in other models and various
modified gravity theories).

For the clarity of the reader, we would like to summarize the main
difference between Ma and Bertschinger's paper \citep{Ma:1995ey} and
our paper. Ma and Bertschinger's starting equations, i.e.,\ Equations\
(29--30) (MB29--30) of \citep{Ma:1995ey}, are
compatible with ours, but are not ready to be implemented for baryons
in Boltzmann codes. In fact, the equations that instead have been
implemented in existing Boltzmann codes are different from ours and
consequently incompatible with MB29--30. This is because the equations
of motion used in existing Boltzmann codes, which correspond instead
to Equations\ (66--67) of \citep{Ma:1995ey} (MB66--67),
have some terms removed from the correct equations, which make them
incompatible with MB29--30. In our code, we have instead implemented
the correct equations of motion that are compatible with MB29--30\footnote{{We provide our new code in the link} \texttt
{http://www2.yukawa.kyoto-u.ac.jp/\textasciitilde antonio.defelice/new\_baryon.zip.}}.

Since the baryon equations of motion MB66--67 (the ones used in Boltzmann
codes) are not compatible with MB29--30 (which instead follow from
conservation of energy-momentum), they end up breaking general
covariance. Therefore, MB66--67 equations describe a system, baryons,
which does not satisfy the energy-momentum conservation.

In the following sections, we also argue that some of the terms removed
from MB29--30 are not a priori small enough before dust domination
(i.e.,\ during the times when the pressure of the baryon cannot be neglected)
and should not be ignored).

Furthermore, manipulating equations MB66--67, instead of MB29--30,
makes this non-conservation of energy-momentum error propagate in
other contexts, especially when we discuss the tight coupling approximation.

As a result, for the first time, we are able to fix energy-momentum
conservation in the baryon sector in Boltzmann codes and, at the same
time, improve the validity of the tight coupling approximation
scheme.

The fact that the equations of motion implemented for the baryon in
existing Boltzmann codes do not satisfy the stress-energy-momentum
conservation (and general covariance) is due to an approximation (which
modified equations MB29--30 into MB66--67) that we want to discuss here.
Inside MB66--67, all the terms that deal with baryonic pressure have
been ignored except for the term $c_{s}^{2}k^{2}\delta_{b}$. Doing
this, once more, leads to breaking of both general covariance and
the conservation of energy-momentum for baryons.

Even if we were to agree with this covariance breaking approximation
scheme, we are at least forced to conclude that whenever there is
a combination $c_{s}^{2}k^{2}\times\left(\text{perturbation field}\right)$;
in general, such a term cannot be ignored a priori, as it may give
non-negligible contributions at sufficiently high $k$.

A non-trivial consequence of this fact is the following. In the synchronous
gauge, in Boltzmann codes, the metric perturbation field $\dot{h}$,
is replaced on using the perturbed $G^{0}{}_{0}$ Einstein equation.
Such an equation replaces everywhere $\dot{h}$ with other terms including
a term proportional to $k^{2}\eta$. On the other hand, one of the
terms that is neglected in the baryon equations of motion MB29--30
is exactly $c_{s}^{2}\dot{h}$. However, by removing this term, one will
end up neglecting a term proportional to $c_{s}^{2}k^{2}\eta$. This
term is absent only if we use the equations MB66--67 (but not with
MB29--30)\footnote{Here, we have used the conventions of Ma and Bertschinger for the metric
perturbation fields $\eta$, $h$.}. While this term during the matter dominated epoch is of the same
order as $a^{2}H^{2}c_{s}^{2}\delta_{b}$ (the Poisson equation would
set $k^{2}\eta\simeq a^{2}H^{2}\delta_{b}$), it can be comparable
to and even larger than $c_{s}^{2}k^{2}\delta_{b}$ during the radiation
dominated epoch. Therefore, as far as we keep $c_{s}^{2}k^{2}\delta_{b}$,
there is no a priori reason why one can ignore $c_{s}^{2}k^{2}\eta$.

In order to avoid all these problems, for us instead, in this paper,
baryons are represented by a non-relativistic fluid for which $T\ll\mu_{g}$
(i.e.,\ $c_{s}^{2}\ll1$), and as such, we do not neglect any term
up to first order in $c_{s}^{2}$. This is the reason why we call
our equation for baryons exact (up to $c_{s}^{2}$ order). Since
we are not ignoring any term, our equation obeys general covariance
and gauge compatibility (once again, up to $c_{s}^{2}$).

While in the present paper, we restrict our consideration to the $\Lambda$CDM
model, our baryon equations can be easily implemented in other models
and various modified gravity theories.

This paper is organized as follows. In Section \ref{BarEqMo}, we
study a baryon fluid from a Lagrangian and derive the equation of
motion in the non-relativistic limit. Then, in Section \ref{tca},
the tight coupling approximation up to the second order is discussed
to overcome the stiffness problem for the new equations of motion
of the coupled baryon-photon system. Subsequently, in Section \ref{CovProb},
we discuss the key problems in the current Boltzmann codes and compare
with our implementation. A brief code implementation is discussed
in Section \ref{CodeImple}. Subsequently, we present the results
of cosmological parameters after doing MCMC analysis in Section \ref{Result}.
Finally, we conclude in Section \ref{Conclusion}.

\section{Baryon Equation of Motion\label{BarEqMo}}

We assume that baryons are described by a non-relativistic fluid.
In that case, we study the dynamics of the perturbations of an ideal
gas, which is meant to represent a more realistic and covariant model
for non-relativistic baryons at early times. For this goal, we introduce
here an action for perfect fluids that is able to describe
the scalar modes of a non-barotropic perfect fluid completely (a class to which
an ideal gas belongs). The vector and tensor modes equations of motion
implemented in Boltzmann codes do not require any corrections, so
that we will only discuss here scalar modes. The needed action can
be written as follows: 
\begin{equation}
S_{m}=-\int d^{4}x\,\sqrt{-g}[\rho(n,s)+J^{\mu}\,(\partial_{\mu}\phi+\vartheta\,\partial_{\mu}s)]\,,\label{eq:Lagr-ideal-gas}
\end{equation}
where $\rho$ represents the fluid energy density, $n$ its number
density, and $s$ its entropy per particle. The other fundamental
variables are the timelike vector $J^{\alpha}$, the metric $g_{\mu\nu}$,
and the scalars $\phi$, $\vartheta$, $s$, whereas: 
\begin{equation}
n\equiv\sqrt{-J^{\mu}J^{\nu}g_{\mu\nu}}\,,\label{eq:def_n}
\end{equation}
and here, since the fluid is non-barotropic, we have $\rho=\rho(n,s)$.
Notice here that the minus sign in Equation~(\ref{eq:def_n}) is needed,
as $J^{\mu}$ represents a timelike vector.

We still need to provide an equation of state, for the perfect fluid,
because we have defined $\rho\equiv\rho\left(n,s\right)$. Here, we
consider a monoatomic ideal gas (a perfect fluid with non-zero, but
small temperature, compared to the particle fluid mass) defined by
the following two equations of state: 
\begin{equation}
p=n\,T\,,\qquad\rho=\mu_{g}\,n+\frac{3}{2}\,n\,T\,,\label{eq:eos-1}
\end{equation}
where $p$ and $T$ are the pressure and the temperature of the fluid,
respectively. Furthermore, $\mu_{g}$ is a constant and represents
the mass of the fluid particle. Since the fluid is assumed to be non-relativistic,
these equations of state hold only for $T\ll\mu_{g}$.

The first law of thermodynamics for a general perfect fluid (see, e.g.,\ \cite{Misner:1974qy})
can be written as: 
\begin{equation}
\text{d}\rho=\mu\,\text{d}n+n\,T\,\text{d}s\,,
\end{equation}
where the enthalpy per particle $\mu$ is defined as:
\begin{equation}
\mu=\frac{\rho+p}{n}\,.
\end{equation}

Although it is well known that these equations of state define an
ideal gas, that is a non-relativistic fluid, we will discuss further
the motivation and the implications of such a choice in Appendix \ref{IdealGas}. 

This action written in these variables, to the best of our knowledge,
has not been introduced before. However, on redefining the vector
variable $J^{\alpha}$ in terms of a vector density, as in $J^{\alpha}=\bar{J}^{\alpha}/\sqrt{-g}$,
the action reduces to the Schutz--Sorkin action of \citep{Schutz:1977df}
(see also, e.g.,\ \citep{Brown:1992kc,DeFelice:2009bx}). The point
of introducing the Lagrangian defined in Equation\ (\ref{eq:Lagr-ideal-gas})
is that it allows us to study general FLRW cosmology with curvature
terms in terms of fields whose interpretation and dynamics are at the
same time simpler and clearer.

The equation of motion for the field $\phi$ gives: 
\begin{equation}
\nabla_{\mu}J^{\mu}=0\,,\label{eq:Eq_phi}
\end{equation}
which is related to the conservation of the number of particles. In fact,
on defining $u^{\mu}$ so that: 
\begin{equation}
J^{\mu}=\,n\,u^{\mu}\,,\label{eq:def_u}
\end{equation}
then Equation~(\ref{eq:def_n}) leads to the following constraint on
$u^{\mu}$: 
\begin{equation}
u^{\mu}u_{\mu}=-1\,.
\end{equation}

Therefore, the timelike vector $u^{\alpha}$ represents the four-velocity
of the fluid, and $J^{\alpha}$ is such a four-velocity vector multiplied
by the four-scalar $n$, the gas number density. This property, in particular,
implies that on a general FLRW background, $J^{i}=0$.

The equation of motion for the field $\vartheta$ leads instead to
the conservation of entropy, namely: 
\begin{equation}
u^{\mu}\nabla_{\mu}s=u^{\mu}\partial_{\mu}s=0\,.\label{eq:conserv_s}
\end{equation}

The equation of motion for the field $s$ leads to $u^{\mu}\partial_{\mu}\vartheta=T$
(because $(\partial\rho/\partial s)_{n}=nT$, from the first law of
thermodynamics), whereas the equations of motion for $J^{\mu}$ relates
$J^{\mu}$ (or $u^{\mu}$) to the other fields $\phi$, $\vartheta$,
$s$.

We can decompose the scalar contributions from the matter field action,
at linear order in perturbation theory about a general FLRW background,
as follows: 
\begin{align}
s & =s_{0}+\delta s\,Y\,,\\
\phi & =-\int^{\tau}d\eta\,a(\eta)\,\overline{\rho}_{,n}+\delta\phi\,Y\,,\\
\vartheta & =\int^{\tau}d\eta\,a^{4}\,\overline{\rho}_{,s}/N_{0}+\delta\vartheta\,Y\,,\\
J^{0} & =\frac{N_{0}}{a^{4}}(1+W_{0}\,Y)\,,\\
J^{i} & =\frac{W}{a^{2}}\,\gamma^{ij}Y_{|j}\,,
\end{align}
where $N_{0}$ represents the total number of fluid particles, and
we have also defined: 
\begin{equation}
\text{d}s^{2}=-(1+2\alpha Y)\,a^{2}\,d\tau^{2}+2a\chi\,Y_{|i}\,d\tau\,dx^{i}+[a^{2}(1+2\zeta Y)\gamma_{ij}+2E\,Y_{|ij}]\,dx^{i}dx^{j}\,.
\end{equation}

Here, $a=a(\tau)$ is the scale factor, $\gamma_{ij}$ is the metric
of a three-dimensional constant-curvature space, the time-independent
function $Y$ is determined by the property $\gamma^{ij}Y_{|ij}=-k^{2}Y$,
and the subscript $|i$ represents the spatial covariant derivative
compatible with the 3D metric $\gamma_{ij}$. All the coefficients
($\delta s$, etc.) are functions of time only.

On an FLRW background, Equation\ (\ref{eq:conserv_s}) leads to $s={\rm {\rm constant}}=s_{0}$.
As a consequence, the perturbation of entropy per particle $\delta s$
becomes gauge invariant and corresponds to a non-adiabatic mode.
On perturbing Equation~(\ref{eq:conserv_s}) at the first order, we find:
\begin{equation}
\delta u^{\mu}\partial_{\mu}s_{0}+u^{0}\partial_{0}\delta s=0\,,
\end{equation}
which, in the real space, implies that: 
\begin{equation}
\frac{\partial}{\partial\tau}\,\delta s=0\,.
\end{equation}
This condition of adiabaticity, i.e.,\ being constant, the entropy 
per baryon is a direct consequence of the equations of state and conservation
of the stress-energy tensor, and it was also correctly considered in Ma
and Bertschinger (see the statement before their Equation\
(96)). Only interactions with photons may affect this, as these may
also affect the exchange of heat. However, quantum mechanically, the
fluids only exchange relative-momentum (which is not zero in general,
even without interactions) at the tree level and, in particular, do not
generate momentum. Therefore, we do not expect the generation of entropy either.
Actually, we can (and have to) choose the initial conditions for the
entropy perturbations (at the end of inflation), so that we have an
adiabatic fluid, namely $\delta s(\vec{x})=0$, having assumed that,
at the end of inflation, no non-adiabatic mode was produced or present.
In this case, since we impose $\delta s$ to vanish, then, because
of this boundary condition, we find $\delta p=c_{s}^{2}\delta\rho+(\partial p/\partial s)_{\rho}\,\delta s=c_{s}^{2}\delta\rho$.

Then, we further find it convenient to define new perturbation variables
$v$, $\delta$, $\delta\vartheta_{v}$, and $\theta$ so that: 
\begin{align}
\delta\phi & =\rho_{,n}\,v-\vartheta(\tau)\,\delta s\,,\\
\delta\vartheta & =\delta\vartheta_{v}-\frac{\rho_{,s}}{n}\,v\,,\\
W_{0} & =\frac{\rho}{n\rho_{,n}}\,\delta-\alpha-\frac{\rho_{,s}}{n\rho_{,n}}\,\delta s\,,\label{eq:W0_delta}\\
v & =-\frac{a}{k^{2}}\,\theta\,,
\end{align}
where Equation~(\ref{eq:W0_delta}) has been found on considering the
definition of the variable $\delta$, namely $\delta\equiv\delta\rho/\bar{\rho}$.
As a result, the two main equations of motion coming from the matter
Lagrangian can be written as: 
\begin{align}
\dot{\theta} & =-\frac{\dot{a}}{a}\,\theta+k^{2}\alpha+\frac{\rho}{\rho+p}\,c_{s}^{2}k^{2}\left(\delta+3\,\frac{\dot{a}}{a}\,\frac{\rho+p}{\rho}\,\frac{\theta}{k^{2}}\right),\label{eq:theta}\\
\frac{\partial}{\partial\tau}\left(\frac{\rho}{\rho+p}\,\delta\right) & =-\theta-3\dot{\zeta}+k^{2}\,\frac{\partial}{\partial\tau}\left(\frac{E}{a^{2}}\right)-\frac{k^{2}}{a}\,\chi\,,
\end{align}
where we have not fixed any gauge yet. Of course, the same equations
can be derived from conservation of the stress-energy tensor, namely
$T^{\mu}{}_{\nu;\mu}=0$.

It can be noticed that, in the first equation, a gauge-invariant combination
associated with $\delta$ is present, namely the comoving matter energy
density perturbation: 
\begin{equation}
\delta_{v}=\delta+3\,\frac{\dot{a}}{a}\,\frac{\rho+p}{\rho}\,\frac{\theta}{k^{2}}\,,
\end{equation}
whereas, in the second equation, another gauge-invariant combination
associated again with $\delta$ appears, namely the flat-gauge energy
density perturbation, or: 
\begin{equation}
\delta_{{\rm FG}}=\delta+\frac{3\,(\rho+p)}{\rho}\,\zeta\,.
\end{equation}

In fact, the second equation can also be rewritten as: 
\begin{equation}
\dot{\delta}=-\frac{\rho+p}{\rho}\left[\theta+3\dot{\zeta}+\frac{k^{2}}{a}\,\chi-k^{2}\,\frac{\partial}{\partial\tau}\left(\frac{E}{a^{2}}\right)\right]-3\frac{\dot{a}}{a}\left[c_{s}^{2}-\frac{p}{\rho}\right]\delta\,.\label{eq:delta}
\end{equation}
where $c_{s}^{2}=\dot{p}/\dot{\rho}=n\rho_{,nn}/\rho_{,n}$ is the
speed of propagation for the fluid (See Appendix \ref{fluids} for detailed discussions about the speed of propagation of matter fields). Equation~(\ref{eq:theta}) and Equation~(\ref{eq:delta})
are the equations of motion for the fluid that we will consider from
now on (in the Subsection \ref{DTcons}, we give an alternative derivation 
of (\ref{eq:theta}) and (\ref{eq:delta}), based on the conservation
of the stress-energy tensor).

In the above derivation the Einstein equation is not used. Therefore, for 
hold for each perfect fluid component, which is separately conserved
and adiabatic, provided that $\rho$, $p$, and perturbation variables
are replaced with the corresponding quantities for the component.

\subsection{Expansion in $T/\mu_{g}$}

Up to now, Equations~(\ref{eq:theta}) and~(\ref{eq:delta}) still hold
for a general adiabatic perfect fluid, i.e.,\ not only for an ideal
gas. From now on, we will instead restrict our consideration to the
case of an ideal gas with a non-zero collision term with a photon
gas. We will fix the equations of motion by making an expansion in
$T/\mu_{g}\ll1$, as the baryon particles are supposed to be non-relativistic.
The dynamical equation for $T$, Equation\ (\ref{eq:Tdot}), reads (see
Appendix \ref{IdealGas} for further details): 
\[
\frac{1}{aH}\frac{\text{d}}{\text{d}\tau}\left(\frac{T}{\mu_{g}}\right)=-2\,\frac{T}{\mu_{g}}+\frac{8}{3}\,\frac{\rho_{\gamma}}{\rho}\,\frac{n_{e}\sigma_{T}}{H}\,\frac{T_{\gamma}-T}{m_{e}}\,,
\]
where we have the Hubble expansion rate as $H=\dot{a}/a^{2}$ (the overdot
denotes a derivative with respect to the conformal time), so that
we will need to assume also that: 
\begin{equation}
\frac{\rho_{\gamma}}{\rho}\,\frac{n_{e}\sigma_{T}}{H}\,\frac{T_{\gamma}-T}{m_{e}}\ll1\,.
\end{equation}

Besides, we have Equations~(\ref{eq:eos}), which imply: 
\begin{align}
c_{s}^{2} & =\left(\frac{\partial p}{\partial\rho}\right)_{s}=\frac{10\,T}{3\,(5T+2\mu_{g})}\approx\frac{5}{3}\,\frac{T}{\mu_{g}}\,.
\end{align}

The Boltzmann equations, on introducing also the interaction term,
can be written as: 
\begin{align}
\dot{\theta} & =-\frac{\dot{a}}{a}\,\theta+k^{2}\alpha+\frac{\rho}{\rho+p}\,c_{s}^{2}k^{2}\left(\delta+3\,\frac{\dot{a}}{a}\,\frac{\rho+p}{\rho}\,\frac{\theta}{k^{2}}\right)+\frac{4\rho_{\gamma}}{3\rho}\,an_{e}\sigma_{T}(\theta_{\gamma}-\theta),\\
\dot{\delta} & =-\frac{\rho+p}{\rho}\left[\theta+3\dot{\zeta}+\frac{k^{2}}{a}\,\chi-k^{2}\,\frac{\partial}{\partial\tau}\left(\frac{E}{a^{2}}\right)\right]-3\,\frac{\dot{a}}{a}\left[c_{s}^{2}-\frac{p}{\rho}\right]\delta\,.
\end{align}

Then, at the lowest order in $T/\mu_{g}$, we find: 
\begin{align}
\dot{\theta} & =-\frac{\dot{a}}{a}\,\theta+k^{2}\alpha\,,\\
\dot{\delta} & =-\theta-3\dot{\zeta}+k^{2}\,\frac{\partial}{\partial\tau}\left(\frac{E}{a^{2}}\right)-\frac{k^{2}}{a}\,\chi\,,
\end{align}
which represents the equations of motion of a cold dust component
with no interactions.

At the first order in $c_{s}^{2}\simeq T/\mu_{g}$, these equations
of motion can be rewritten so that they look as similar as possible
to the ones present in \citep{Ma:1995ey}, as follows: 
\begin{align}
\dot{\theta}_{b} & =-\frac{\dot{a}}{a}\,\theta_{b}+k^{2}\alpha+c_{s}^{2}\,k^{2}\left(\delta_{b}+3\,\frac{\dot{a}}{a}\,\frac{\theta_{b}}{k^{2}}\right)+R\,an_{e}\sigma_{T}(\theta_{\gamma}-\theta_{b})\,,\label{eq:dottheta}\\
\dot{\delta}_{b} & =-\frac{6}{5}\,c_{s}^{2}\,\frac{\dot{a}}{a}\,\delta_{b}-\left(1+\frac{3}{5}\,c_{s}^{2}\right)\left[\theta_{b}+3\dot{\zeta}-k^{2}\,\frac{\partial}{\partial\tau}\left(\frac{E}{a^{2}}\right)+\frac{k^{2}}{a}\,\chi\right],\label{eq:dotdelb}
\end{align}
where $R\equiv\frac{4}{3}\,\rho_{\gamma}/\rho_{b}$, $c_{s}^{2}$
can be found by solving Equation~(\ref{eq:speed_T}) and the subscripts
$b$ and $\gamma$ denote the baryon and photon, respectively. Both of
these equations of motion for the baryon perturbation variables are
gauge invariant (up to the first order in $c_{s}^{2}$). In case higher
precision is needed, then one can further write down the equations
of motion at any order in $T/\mu_{g}$. In this work, we will only
consider the first order approximation corrections to the dust fluid
case, and we will apply them in a consistent and covariant way to
a well-known Boltzmann code, CLASS.

Actually, the above two equations are equivalent to MB29--30, but indeed
differ from the baryon equations of motion MB66--67 given in \citep{Ma:1995ey}:
they have to be, as the latter ones are not covariant. In this work,
we claim that, on introducing these covariant equations, we can solve
all the problems we have already stated in the introduction. The solution 
here merely comes from the fact that our baryon equations of motion
have been derived directly from a covariant action, and on top of
that, we are expanding them in terms of $c_{s}^{2}=(\partial p/\partial\rho)_{s}$,
which is a scalar.

\subsection{Baryon Covariant Equations of Motion from the Conservation Law\label{DTcons}}

In this subsection, we show that the equations of motion for a non-barotropic
perfect fluid obtained from the action can be also derived using a
different approach, namely from the conservation law $\nabla_{\mu}T^{\mu}{}_{\nu}=0.$

Let us suppose we have a perfect fluid in the conventional form: 
\begin{equation}
T^{\mu}{}_{\nu}=\left(\rho+p\right)u^{\mu}\,u_{\nu}+p\,\delta^{\mu}{}_{\nu},
\end{equation}
where $u^{\mu}$ is the four-velocity of the perfect fluid. We have
the usual constraint on the fluid velocity: 
\begin{equation}
u_{\mu}u^{\mu}=-1.
\end{equation}

The linear perturbations are defined as: 
\begin{align}
\rho & =\bar{\rho}+\delta\rho\,,\\
p & =\bar{p}+\delta p\,,\\
u_{0} & =-a\left(1+\delta u\right)\,,\\
u_{i} & =\partial_{i}v_{s}\,,\\
g_{00} & =-a^{2}\left(1+2\alpha\right)\,,\\
g_{0i} & =a\partial_{i}\chi\,,\\
g_{ij} & =a^{2}\left[\left(1+2\zeta\right)\delta_{ij}+\frac{2\partial_{i}\partial_{j}E}{a^{2}}\,\right]\,.
\end{align}
where $\delta u$ (which is determined by the condition $u_{\mu}u^{\mu}=-1$)
and $v_{s}$ are the scalar perturbations in the fluid velocity. Since
$p$ can be written in terms of two other thermodynamical variables,
pressure perturbation $\delta p=\left(\partial p/\partial\rho\right)_{s}\delta\rho\,+\,\left(\partial p/\partial\rho\right)_{\rho}\delta s$,
where $\left(\partial p/\partial\rho\right)_{s}\equiv c_{s}^{2}$.
Considering the adiabatic initial condition, we can choose $\delta s=0$.
Hence, $\delta p=\left(\partial p/\partial\rho\right)_{s}\delta\rho=c_{s}^{2}\delta\rho$.
Notice that we have not chosen any gauge. Now, we have the component
of energy-momentum tensor as: 
\begin{align}
T^{0}{}{_{0}} & =-\bar{\rho}\left(1+\delta\right)\,,\\
T^{0}{}{_{i}} & =\left(\bar{\rho}+\bar{p}\right)\frac{\partial_{i}v_{s}}{a}\,,\\
T^{i}{}{_{j}} & =\left(\bar{p}+\delta p\right)\delta^{i}{}{_{j}}\,,
\end{align}
where $\delta\equiv\delta\rho/\rho$. Here, for simplicity, we do
not consider anisotropic shear perturbations. Now, to find the equations
of motion, the calculation is straight forward. From the conservation 
law: 
\begin{equation}
\nabla_{\mu}T^{\mu\nu}=\partial_{\mu}T^{\mu\nu}+\Gamma^{\mu}\,_{\mu\beta}\,T^{\beta\nu}+\Gamma^{\nu}\,_{\mu\beta}\,T^{\mu\beta}=0,
\end{equation}
we find the equations of motion for linear perturbation in the Fourier
space: 
\begin{align}
\dot{\delta} & =-\frac{\rho+p}{\rho}\left[\theta+3\dot{\zeta}+\frac{k^{2}\chi}{a}-k^{2}\frac{\partial}{\partial t}\left(\frac{E}{a^{2}}\right)\right]-3\frac{\dot{a}}{a}\left(c_{s}^{2}-\frac{p}{\rho}\right)\delta\,,\\
\dot{\theta} & =-\frac{\dot{a}}{a}\theta+k^{2}\alpha+c_{s}^{2}\left(3\frac{\dot{a}}{a}\theta+k^{2}\frac{\rho}{\rho+p}\delta\right)\,,
\end{align}
where we have redefined $v_{s}=-\theta a/k^{2}$ and $c_{s}^{2}=(\partial p/\partial\rho)_{s}$
evaluated at the background level. The above two equations of motion
exactly match Equation\
(\ref{eq:delta}) and Equation\ (\ref{eq:theta}), respectively, which
are the resulting equations from the variation of the action (\ref{eq:Lagr-ideal-gas}).
This ensures that the action discussed in Section \ref{BarEqMo}
is a well defined covariant action for a perfect fluid. These equations
are equivalent to MB29--30 equations of motion, after we implement
the equations of state for the fluid.

We believe that these are the equations of motions that need to be
implemented in any Boltzmann code; otherwise, baryon physics will be
described out of general relativity.

\section{Tight Coupling Approximation\label{tca}}

In 1970, Peebles and Yu \citep{Peebles:1970ag} introduced a technique
to solve the cosmological evolution of a tightly coupled photon-baryon
fluid. The interaction time scale of photons and baryons is given
by $\tau_{c}\equiv\left(an_{e}\sigma_{T}\right)^{-1}$, where $\sigma_{T}$
is the Thomson scattering amplitude. This time scale of the interaction
is shorter than both sub-horizon and super-horizon scales, on which
most of the modes of our interest are evolved. At the time when photons
and baryons are tightly coupled together, the dynamical equations
of motion become stiff, so that standard numerical integrators become
invalid. They solved this system perturbatively in $\tau_{c}$ for
terms that are considerably small in the limit $\tau_{c}\rightarrow0$.
These perturbative solutions are implemented numerically in the Boltzmann
code. Here, we recalculate the tight coupling approximation equations
using the gauge invariant equation of motion of baryons derived in
the previous section.

The first order tight coupling approximation was implemented in \citep{Ma:1995ey}
by making two additional assumptions/approximations, namely $\tau_{c}\propto a^{2}$
and $c_{s}^{2}\propto a^{-1}$. CAMB \citep{Lewis:1999bs} also has
implemented the first order approximation, assuming only $\tau_{c}\propto a^{2}$.
Here, we show our results of tight coupling approximation, with the
covariance obeying the equations of motion for baryons. The detailed calculations
up to second order are given in Appendix \ref{tcacal}.

At first order in tight coupling approximation, we have:

\begin{equation}
\dot{\Theta}_{\gamma b}=\left(\frac{\dot{\tau}_{c}}{\tau_{c}}-\frac{2\mathcal{H}}{R+1}\right)\Theta_{\gamma b}+\mathcal{T}_{1}+\mathcal{O}(\tau_{c}^{2})\,,
\end{equation}
where: 
\begin{align}
\mathcal{T}_{1} & \equiv\frac{\tau_{c}}{R+1}\left[\left(\mathcal{H}^{2}+\dot{\mathcal{H}}\right)\theta_{b}+\left(\frac{\dot{\delta}_{\gamma}}{4}-c_{s}^{2}\dot{\delta}_{b}+\mathcal{H}\alpha+\frac{\mathcal{H}\delta_{\gamma}}{2}-\bar{c}_{s}^{2}\delta_{b}\right)k^{2}\right]\nonumber \\
 & -\frac{3\tau_{c}}{R+1}\left[\frac{\left\{ 3\mathcal{H}^{2}c_{s}^{4}+[\dot{\mathcal{H}}\left(R+1\right)-\mathcal{H}^{2}]c_{s}^{2}+\mathcal{H}\bar{c}_{s}^{2}\left(R+1\right)\right\} \theta_{b}}{R+1}+\mathcal{H}c_{s}^{2}k^{2}\!\left(\alpha+\frac{1}{4}\,\frac{R\delta_{\gamma}}{R+1}+\frac{c_{s}^{2}\delta_{b}}{R+1}\right)\right]\,.\label{eq:slip_T1}
\end{align}

Comparing our results to the ones in \citep{Blas:2011rf}, the new
parts of this equation consist of the following two parts: (1) in the
$\alpha$ term in the first line (corresponding to the fact that a
gauge has not been chosen yet) and (2) in the entire second line, which
is due to gauge choice plus the corrections to the baryon dynamics.

In this last equation as well, even if we consider the high $k$ limit,
we should not neglect the new terms proportional to either $\alpha$
or $\delta_{\gamma}$. We stress once more here that the new terms
we get appear only when we use the correct equations MB29--30. On using
instead MB66--67, we would miss these terms all at once. In particular,
as we will show in the results (compare Figure\ \ref{fig:perturbation_plot}),
we cannot neglect the term proportional to $c_{s}^{2}k^{2}\delta_{\gamma}\,R/(R+1)\approx c_{s}^{2}k^{2}\delta_{\gamma}$.

\section{Comparison between Current Boltzmann Codes and Covariance Full Equations
\label{CovProb}}

In this section, we want to show explicitly that MB66--67 equations
of motion for the baryons do break general covariance. This should
help also the potential problem for Boltzmann solvers, which implement
these equation. The results would depend on the gauge, if general covariance
is broken inside the equations of motion.

Let us focus our attention on the scalar perturbations of a {flat Friedmann--Lema\^{i}tre--Robertson--Walker}
(FLRW) metric, which can be written as follows: 
\begin{equation}
\text{d}s^{2}=-\left(1+2\alpha\right)a^{2}\,d\tau^{2}+2a\partial_{i}\chi\,d\tau\,dx^{i}+a^{2}\left[\left(1+2\zeta\right)\delta_{ij}+\frac{2\partial_{i}\partial_{j}E}{a^{2}}\,\right]dx^{i}dx^{j}\,.
\end{equation}

In the above, we have not fixed any gauge yet.

Then, on choosing the Newtonian gauge, i.e.,\ on setting $\chi=0$
and $E=0$, the equations of motion for the baryon fluid MB29--30 as
given in \citep{Ma:1995ey} (their Equation~(67)), which are also used
in current Boltzmann codes, read in the Fourier space as follows:
\begin{align}
\dot{\delta}_{b} & =-\theta_{b}-3\dot{\zeta},\label{eq:Mabaryondelta}\\
\dot{\theta}_{b} & =-\frac{\dot{a}}{a}\theta_{b}+k^{2}\alpha+c_{s}^{2}k^{2}\delta_{b}+\frac{4\bar{\rho}_{\gamma}}{3\bar{\rho}_{b}}\,an_{e}\sigma_{T}\,(\theta_{\gamma}-\theta_{b}),\label{eq:MaBaryontheta}
\end{align}
where $\delta_{b}\equiv\left(\rho_{b}-\bar{\rho}_{b}\right)/\bar{\rho}_{b}$
is the baryon density perturbation, $\theta_{b}$ is the scalar part
of the baryon velocity perturbation defined as $\left(\bar{\rho}_{b}+\bar{P}_{b}\right)\theta_{b}\equiv ik^{j}\delta T_{b}^{0}\,_{j}$,
$\theta_{\gamma}$ is the scalar part of the photon velocity perturbation,
and the term $an_{e}\sigma_{T}\,(\theta_{\gamma}-\theta_{b})$ represents
the momentum transfer into the photon gas. A dot here represents a
derivative with respect to $\tau$, the conformal time. Equations
(\ref{eq:Mabaryondelta}) and (\ref{eq:MaBaryontheta}) are written
in the Newtonian gauge, but can be rewritten in terms of the gauge
invariant variables (which reduce to the corresponding perturbation
variables in the Newtonian gauge when $\chi=E=0$): 
\begin{align}
\delta_{b}^{{\rm GI}} & =\delta_{b}+\frac{\dot{\bar{\rho}}_{b}}{a\bar{\rho}_{b}}\chi-\frac{\dot{\bar{\rho}}_{b}}{\bar{\rho}_{b}}\frac{\partial}{\partial\tau}\left(\frac{E}{a^{2}}\right),\\
\theta_{b}^{{\rm GI}} & =\theta_{b}+\frac{k^{2}}{a}\chi-k^{2}\frac{\partial}{\partial\tau}\left(\frac{E}{a^{2}}\right),\\
\theta_{\gamma}^{{\rm GI}} & =\theta_{\gamma}+\frac{k^{2}}{a}\chi-k^{2}\frac{\partial}{\partial\tau}\left(\frac{E}{a^{2}}\right),\\
\zeta^{{\rm GI}} & =\zeta+\frac{\dot{a}}{a^{2}}\chi-\frac{\dot{a}}{a}\frac{\partial}{\partial\tau}\left(\frac{E}{a^{2}}\right),\\
\alpha^{{\rm GI}} & =\alpha+\frac{1}{a}\dot{\chi}-\frac{\dot{a}}{a}\frac{\partial}{\partial\tau}\left(\frac{E}{a^{2}}\right)-\frac{\partial^{2}}{\partial\tau^{2}}\left(\frac{E}{a^{2}}\right),
\end{align}
as: 
\begin{align}
\dot{\delta}_{b}^{{\rm GI}} & =-\theta_{b}^{{\rm GI}}-3\dot{\zeta}^{{\rm GI}},\label{eq:Mabaryondelta-1}\\
\dot{\theta}_{b}^{{\rm GI}} & =-\frac{\dot{a}}{a}\theta_{b}^{{\rm GI}}+k^{2}\alpha^{{\rm GI}}+c_{s}^{2}k^{2}\delta_{b}^{{\rm GI}}+\frac{4\bar{\rho}_{\gamma}}{3\bar{\rho}_{b}}an_{e}\sigma_{T}\left(\theta_{\gamma}^{{\rm GI}}-\theta_{b}^{{\rm GI}}\right),\label{eq:MaBaryontheta-1}
\end{align}

Since the general covariance is supposed to hold (as we are discussing
general relativity in the presence of standard matter), we are now
able to rewrite the previous evolution equations in any other gauge,
in particular in the synchronous gauge, for which $\alpha=0$ and
$\chi=0$. Then, the dynamical Equations (\ref{eq:Mabaryondelta-1})
and (\ref{eq:MaBaryontheta-1}) reduce to\footnote{\label{classredef_foot}The relation between fields defined in this
paper and definitions used in CLASS are $2E/a^{2}=-1/k^{2}\left(h+6\eta\right)$
and $\zeta=-\eta$. For the synchronous gauge, we add the gauge choice
$\alpha=0$, $\chi=0$, which is actually incomplete. For a complete
gauge fixing, we need to choose also the following two initial conditions
at the time $\tau=\tau_{{\rm ini}}$: $\theta_{c}(\tau_{{\rm ini}})=0$,
$\delta_{\gamma}(\tau_{{\rm ini}})+\frac{2}{3}h(\tau_{{\rm ini}})=0$,
where $\theta_{c}$ represents the field $\theta$ for the cold dark
matter fluid and $\delta_{\gamma}$ is the photon density perturbation.
For the Newtonian gauge, the authors in \citep{Ma:1995ey} used the following
field redefinition, $\alpha=\psi$, $\zeta=-\phi$, together with
the complete gauge fixing, $E=0$, $\chi=0$.}: 
\begin{align}
\dot{\delta}_{b} & =-\theta_{b}+k^{2}\frac{\partial}{\partial\tau}\left(\frac{E}{a^{2}}\right)-3\dot{\zeta}\,,\\
\dot{\theta}_{b} & =-\frac{\dot{a}}{a}\theta_{b}+c_{s}^{2}k^{2}\left[\delta_{b}+3\frac{\dot{a}}{a}\frac{\partial}{\partial\tau}\left(\frac{E}{a^{2}}\right)\right]+\frac{4\bar{\rho}_{\gamma}}{3\bar{\rho}_{b}}an_{e}\sigma_{T}\left(\theta_{\gamma}-\theta_{b}\right),\label{eq:wrong-synchro}
\end{align}
where now the fields are all evaluated in the synchronous gauge. Notice
that the interaction term proportional to $\sigma_{T}$ is gauge invariant
since $\theta_{\gamma}-\theta_{b}=\theta_{\gamma}^{{\rm GI}}-\theta_{b}^{{\rm GI}}$.

Notice that the above differential equation for the velocity field,
Equation~(\ref{eq:wrong-synchro}), is different from the one written
in MB66--67 (precisely Equation~(66) of \citep{Ma:1995ey}), which is also
supposed to be the baryon equation of motion written in the synchronous
gauge. More precisely, the term proportional to $k^{2}c_{s}^{2}$
makes the baryon velocity equation incompatible between the two gauges.
To look at this same problem from another point of view, we can start
by writing the dynamical baryon equations of motion in the synchronous
gauge, as given in Equation~(66) of \citep{Ma:1995ey}, and then transform
them to the Newtonian gauge. However, on doing so, the resulting baryon-velocity
equation of motion turns out to be once again different from the one
shown in Equation~(\ref{eq:MaBaryontheta}) (or Equation~(67) in \citep{Ma:1995ey}).

Therefore, up to now, solving Boltzmann equations in the two gauges
leads to solving two intrinsically different equations of motion,
so that the two gauges give rise to two physically different solutions.
Then, one may wonder which of the two should be considered.

First of all, it is obvious from Equation~(\ref{eq:MaBaryontheta}) that
setting $c_{s}^{2}$ to vanish would make the baryon-velocity equation
compatible among the two gauges; see also the results in \cite{Dodelson:2003ft}
(and \cite{Bartolo:2007ax} to study its extension to second order
perturbation theory).

One should now be convinced that Equations~(\ref{eq:Mabaryondelta}) and
(\ref{eq:MaBaryontheta}) break covariance, but may think this is
due merely to an approximation. Let us then consider {which}
approximation has been considered. We have already stated above that
on top of considering baryons to be non-relativistic, a small scale
limit has been taken, namely $k\gg aH=\dot{a}/a\equiv\mathcal{H}$,
but we will show it here explicitly. As we shall see later on, the
correct equations of motion for the baryon fluid, in the synchronous gauge
(see Footnote \ref{classredef_foot} for the redefinition of field
variables), can be written as (the subscript $b$ indicates the baryon):
\begin{eqnarray}
\dot{\theta}_{b} & = & -\frac{\dot{a}}{a}\,\theta_{b}+c_{s}^{2}\,k^{2}\,\delta_{b}+R\,an_{e}\sigma_{T}(\theta_{\gamma}-\theta_{b})+3\,c_{s}^{2}\,\frac{\dot{a}}{a}\,\theta_{b}\,,\label{eq:dottheta-1b}\\
\dot{\delta}_{b} & = & -\theta_{b}-\frac{1}{2}\,\dot{h}-\frac{6}{5}\,c_{s}^{2}\,\frac{\dot{a}}{a}\,\delta_{b}-\frac{3}{5}\,c_{s}^{2}\left(\theta_{b}+\frac{1}{2}\,\dot{h}\right),\label{eq:dotdelb-1b}
\end{eqnarray}
whereas the non-covariant equations of motion read $\dot{\delta}_{b}=-\theta-\frac{\dot{h}}{2}$
and $\dot{\theta}_{b}=-\frac{\dot{a}}{a}\theta_{b}+c_{s}^{2}k^{2}\delta_{b}$.
As stated in the previous section, in Boltzmann codes, $\dot{h}$
is actually replaced by using the Hamiltonian constraint, which includes
a term $c_{s}^{2}k^{2}\eta$, and so, we cannot remove the $c_{s}^{2}\dot{h}$
term a priori, even following the Ma and Bertschinger approximation choice
for baryons.

Since we only want to know the approximation done and its meaning,
let us switch off the coupling with the photons, take the time
derivative of $\dot{\delta}_{b}$, and represent this expression as
$\ddot{\delta}_{b}=\dot{P}$ where $P$ corresponds to the rhs of
Equation~(\ref{eq:dotdelb-1b}). Then, we can also write ${\ddot{\delta}}_{b}+\mathcal{H}\,{\dot{\delta}}_{b}=\dot{P}+\mathcal{H}\,P$
(where $\mathcal{H}=\dot{a}/a$ and time is conformal). Substituting
in the rhs of such equation both $\dot{\theta}_{b}$ and $\dot{\delta}_{b}$
by the above given equations and replacing $\dot{h}$ and $\ddot{h}$
(also, this latter term introduces terms proportional to $k^{2}\eta$)
by the Einstein equations, we can write the second order baryon equation
in the form\footnote{\label{classredef_foot-1}The extra terms in $c_{s}^{2}$ is due
to the fact that Equation~(\ref{eq:dottheta-1b}) and Equation~(\ref{eq:dotdelb-1b})
have $c_{s}^{2}$ terms that are not neglected a priori. On using
Einstein equations in the synchronous gauge, one gets these extra terms; especially, the term $c_{s}^{2}k^{2}\eta$, which can never be neglected
since $\delta_{b}<\eta$ in the relevant redshift range, as we show
in Figure~\ref{fig:perturbation_plot}.}: 
\begin{eqnarray}
{\ddot{\delta}}_{{b}}+\mathcal{H}{\dot{\delta}}_{b} & = & -\left[{c_{{s}}^{2}}{k}^{2}+1/25\,(120\,{\mathcal{H}}^{2}+30\,{\dot{\mathcal{H}}}){c_{{s}}^{2}}+3\,\mathcal{H}\dot{c_{s}^{2}}\right]\delta_{{b}}\label{eq:lewis_corrected}\\
 & & {}-{\frac{3\,\mathcal{H}{c_{{s}}^{2}}\theta}{5}}+{\frac{12}{5}}\,c_{s}^{2}{k}^{2}\eta+{\frac{12}{5}\left(3\,{c_{{s}}^{2}}+5\right)\pi G_{{N}}a^{2}\,\sum_{i}\delta p_{i}}\nonumber \\
 & & {}+{4(1+3\,{c_{{s}}^{2}})\pi G_{{N}}a^{2}\,\sum_{i}\delta\rho_{i}}\,.\nonumber 
\end{eqnarray}

On comparing this last equation with the one given in \cite{Lewis:2007zh}
for baryon sector, namely:

\begin{equation}
\ddot{\delta}_{b}+\mathcal{H}\dot{\delta}_{b}+c_{s}^{2}k^{2}\delta_{b}=4\pi G_{N}a^{2}\sum_{i}\left(\delta\rho_{i}\,+3\delta p_{i}\right),\label{eq:lewis2007eq}
\end{equation}
we see all the additional terms that have been neglected. As expected,
a term proportional to $c_{s}^{2}k^{2}\eta$ has appeared, and as
we shall see later in Figure\ \ref{fig:perturbation_plot}, it cannot
be neglected a priori during the radiation domination era, up to dust
domination.

Nonetheless, the missing terms correspond to the realization and the
definition of the approximation itself. This approximation consists
of taking $k\gg aH$, $\delta_{b}\gg\eta$ (where $H=\dot{a}/a^{2}$
and time is conformal) and $k^{2}\delta_{b}\gg aH\theta$. However,
during radiation domination (valid at least for the initial conditions
taken in \citep{Ma:1995ey} at $z\geq10^{6}$) at which the speed
of propagation for the baryons cannot be neglected (as $c_{s}^{2}\propto a^{-1}$
approximately), on considering $H\approx H_{0}\sqrt{\Omega_{r0}}(1+z)^{2}$,
where $H_{0}\approx h{\rm Mpc}^{-1}/(2997.9)$, we find that $k$
is constrained to be $k\gg(1+z)\sqrt{\Omega_{r0}}/(3\times10^{3})\,h{\rm Mpc}^{-1}$.
Furthermore, even considering the high $k$ limit, in the equations,
we should at least keep the term proportional to $c_{s}^{2}k^{2}\eta$.

We have another strong reason why one should not ignore the terms
in $c_{s}^{2}$ that instead we have kept. Let us explain this more
in detail. When we study the tight coupling approximation scheme,
on using the approximated baryon equations of motion of \citep{Ma:1995ey},
one is bound to miss some relevant terms (relevant in the sense of the point of view of
\citep{Ma:1995ey}), e.g.,\ $c_{s}^{2}k^{2}\psi$ (in the
Newtonian gauge) and $c_{s}^{2}k^{2}\delta_{\gamma}$ (in any gauge),
which are meant, a priori, to contribute at sufficiently high $k$,
Equation\
(\ref{eq:slip_T1}). Removing these terms (as they do not appear in
previous Boltzmann codes) leads in general to self inconsistencies
inside the regime of the validity of Ma and Bertschinger's approximation.

As we have seen before, in Section \ref{BarEqMo}, introducing back
the gauge compatibility and the general covariance ends up with considering
a new set of equations of motion for the baryon fluid, which will
give different results from the equations found in \citep{Ma:1995ey}.
We have also observed that making the equations of motion explicitly
covariant will not make the numerical code unstable, or slow. Therefore,
we do not have a clear reason why the corrections that are introduced
should not be implemented in today's Boltzmann solvers. It is true
that the corrections are not large enough to change the final results
beyond the error bar, but in the equations of motion, we see that
these corrections (of order of $k^{2}c_{s}^{2}$) are of a similar
order as second order tight coupling approximation quantities. Therefore,
implementing tight coupling approximation correctly for the aim of
reaching precision cosmology should also lead to considering exact and
covariant baryon equations of motion.

We point out here that there is another problem related to the gauge incompatibility
of the equations of motion. In fact, since the equations of motion
are not gauge compatible, we should conclude that the general covariance
has been broken. In turn, this behavior leads to the fact that the
equations of motion will not close in general. That is, the Bianchi
identities will not hold any longer for the system of the perturbed
Einstein equations. If Bianchi identities do not hold any longer,
then in general, picking up a subset of equations will lead to a
solution that does not solve the other remaining equations. This
implies that in general, there is no solution to the full set of equations.

Finally, the new equations of motions introduced here should be considered
as the basic baryon equations of motion. Therefore, any other additional
physical phenomenon that has to do with baryon physics, such as reionization and
higher order tight coupling approximation approximation schemes, should
be considered as starting from the most basic level, i.e.,\ from the
equations of motion that we introduced in this paper.

\begin{figure}[ht]
\includegraphics[width=17cm]{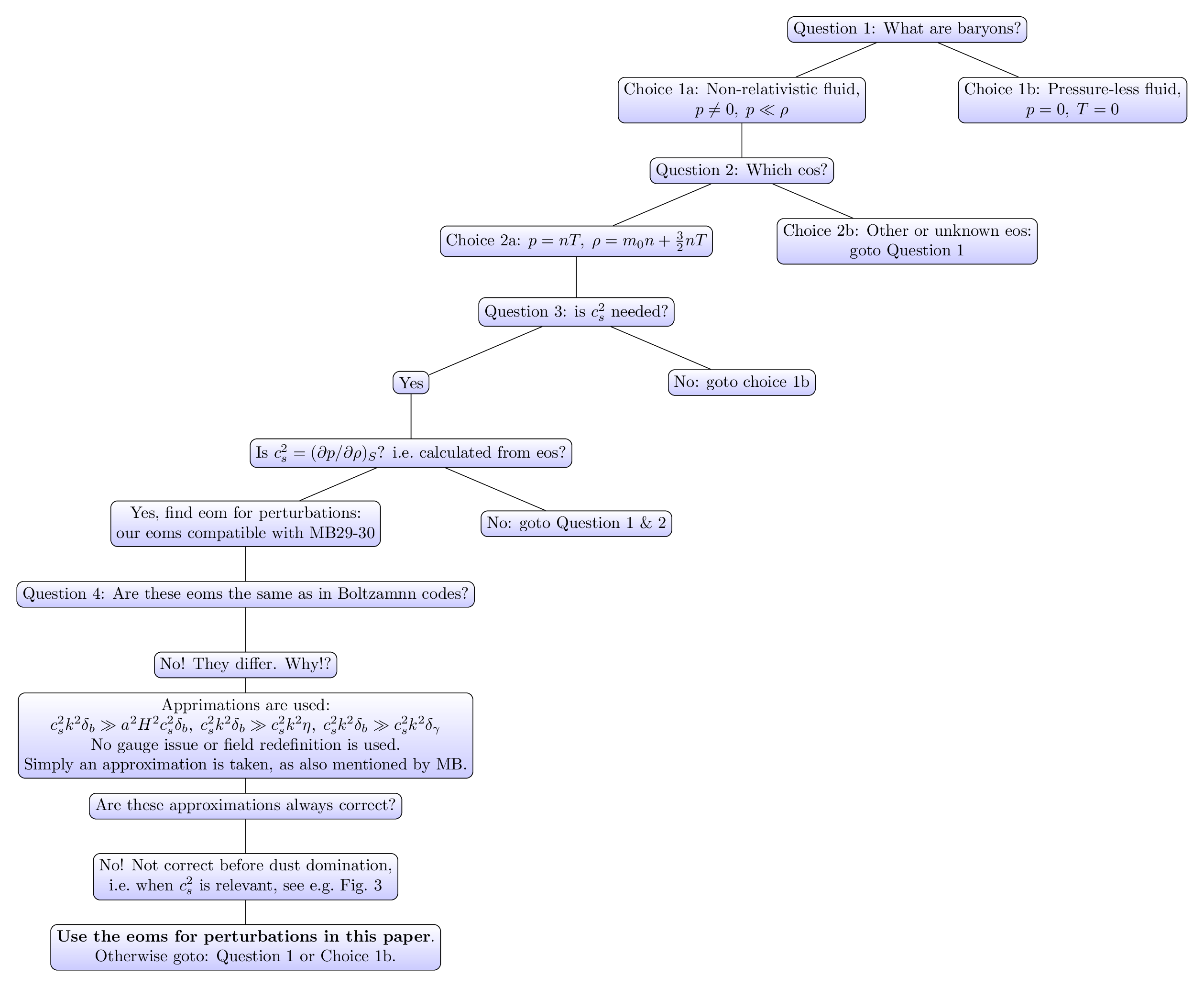} \caption{{Tree diagram that describes the logic followed in this paper to address
the issue of non-conservation of energy-momentum in the equations}
of motion MB66--67 for baryons in existing Boltzmann codes.\label{fig:tree}}
\end{figure}

To summarize here the whole philosophy of our research path,
we introduce the following tree diagram in Figure\ \ref{fig:tree}.

\section{Code Implementation\label{CodeImple}}

We chose the CLASS Boltzmann solver to implement the corrected baryon
and tight coupling approximation equations. We will also make our
notation as close as possible to that used in \citep{Ma:1995ey,Blas:2011rf}.
So far, we have not fixed any gauge, but in the code, we chose the
synchronous gauge for the tight coupling approximation 
scheme. It is straightforward to implement our approximation schemes
in any other gauge.

Since we chose the synchronous gauge, $\alpha=0=\chi$. Furthermore,
we will make the following field redefinitions: 
\begin{align}
-\partial_{\tau}\!\left(\frac{E}{a^{2}}\right) & \equiv\alpha_{{\rm CL}},\\
\zeta & =-\eta_{{\rm CL}}\,,\\
k^{2}\,\frac{E}{a^{2}} & =-\frac{1}{2}\,(h_{{\rm CL}}+6\,\eta_{{\rm CL}})\,,\\
\alpha_{{\rm CL}} & =\frac{1}{2k^{2}}\,(\dot{h}_{{\rm CL}}+6\,\dot{\eta}_{{\rm CL}})\,,
\end{align}

CLASS has implemented five different tight coupling approximation
schemes. In light of the new baryon equations, we also performed
all these approximations except for the one named ``second\_order\_CSR''.
As mentioned above for the { Ma \& Bertschinger }
 linear approximation scheme, we made
the approximations $\tau_{c}\propto a^{2}$ and $c_{s}^{2}\propto a^{-1}$,
for the slip parameter at first order. As for the first order CAMB,
we only considered the approximation $c_{s}^{2}\propto a^{-1}$. For
both first order and second order class schemes, we did not make any
approximation for $\tau_{c}$ or $c_{s}^{2}$.

The parameter estimation using Monte Python was carried out in a super-computer,
XC-40, having 64 nodes, each node having 64 cores. In total, 4096 cores
were available. We chose to run three independent series\footnote{One for each author.}
of 1024 chains, each chain using four cores in parallel. For each chain,
we performed 13,000 steps. We ran all four tight coupling
approximation schemes we implemented and found that they were
all compatible with each other. On running the code, we also
found that all the tight coupling approximation approximation schemes
we implemented did not make the code any slower or stiffer\footnote{The code can be found at $\texttt{http://www2.yukawa.kyoto-u.ac.jp/\textasciitilde antonio.defelice/new\_baryon.zip}$}.
In particular, the time needed by each run to terminate differed
by a few minutes over approximately eight hours. As for stability,
we did not have to increase precision or reduce the time step in order
to solve the perturbation equations of motion. It was also noted that
the acceptance rate for MCMC analysis was increased by about 0.95$\%$
for our improved code implementation with respect to the old code.
This fact usually implies that the system has found a lower minimum
for the $\chi^{2}$. The likelihood of the covariantly corrected code
was slightly improved to $-\ln{\cal L}_{\mathrm{min}}=5984.11$ with
respect to the old covariance breaking code for which the likelihood
was $-\ln{\cal L}_{\mathrm{min}}=5984.45$.

\section{Results\label{Result}}

\begin{figure}[ht]
\includegraphics[width=17cm]{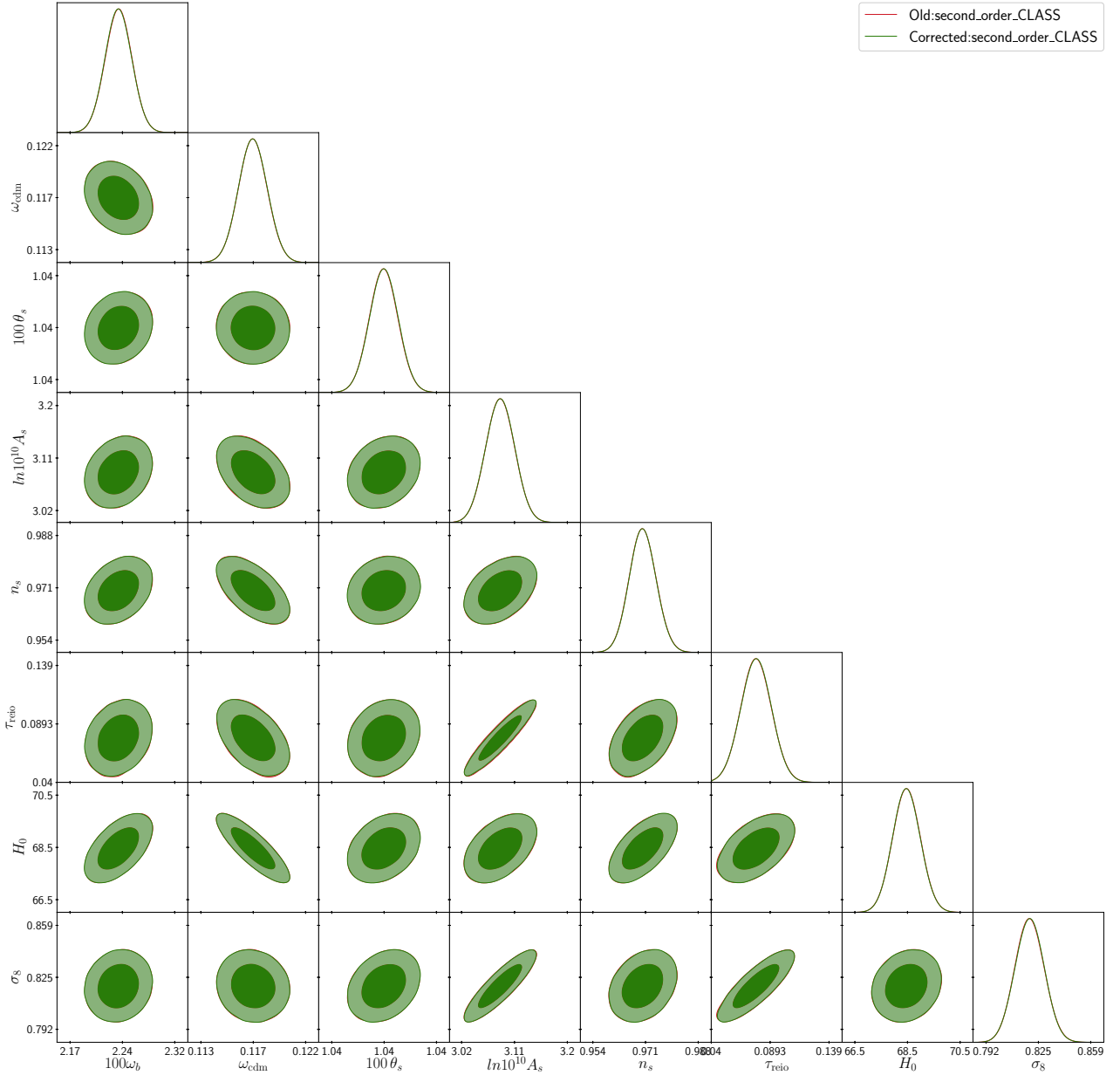} \caption{Combined results given by the old, original CLASS code with covariance breaking
equations of motion for the baryon and the new code with covariant
equations of motion for the baryon fluid. The percent level differences
shown in Table~\ref{tab:Cosmological-Parameters} are invisible in
this figure, but will be important for future surveys.\label{fig:combined}}
\end{figure}

Here, we present our results of running the Monte Carlo sampler for
the cosmological parameter estimation. For analysis, we used the following
datasets: Planck 2015 ({high l}
, {low l,} and lensing) \cite{Adam:2015rua},
{JLA} \cite{Betoule:2014frx}, {BAO}
 {BOSS}
{ DR12}
 \cite{Alam:2016hwk}, BAO
{SMALLZ}
 2014 \cite{2011MNRAS.416.3017B,Ross:2014qpa}, and Hubble Space
Telescope \cite{Riess:2016jrr}. We compared the estimation for the
cosmological parameters between the old baryon equations (which were
non-covariant) and the new covariant ones. We ran the Monte Python
sampler for all four different tight coupling approximation schemes
we implemented in CLASS,{ namely first}\_
order\_Ma, first\_order\_CAMB,
first\_order\_CLASS, and second\_order\_CLASS. Below, in Table~\ref{tab:Cosmological-Parameters},
we show the results for the second order tight coupling approximation
scheme given by the new covariant equations of motion and the results
according to old covariance breaking baryon equations of motion for
the same second order tight coupling approximation scheme. Furthermore, we
give combined plots of the old and the new code for the second order
tight coupling approximation scheme in Figure \ref{fig:combined}. Here,
we only show the numerical results for this scheme, as this is the
one whose code underwent the largest number of modifications.

We found that the new results for the cosmological parameters numerically
agreed with the previous results within one percent, and this fact
was reassuring. Nonetheless, our corrections to the equations of motion
gave a contribution that was not completely negligible, and we believe
this improvement could give a useful contribution in the context of
``precision cosmology'' \citep{Ade:2018sbj,Lazear_2014}.

The amount of changes in the estimation of the parameters was similar
among the different tight coupling approximation schemes we implemented.
In fact, the magnitude of such changes we obtained was of the
same size of the difference in the results that the original non-covariant
code was giving for the different tight coupling approximation schemes.
This means that in order to address the needed precision in the context
of the newest (and future) cosmological probes, we should use Boltzmann
solvers with the corrected covariant baryon dynamical equations.

\begin{table}
\caption{Best fit values of cosmological parameters given by MCMC analysis:
old vs. new. The upper and lower limits are at the 95\% confidence level.\label{tab:Cosmological-Parameters}}

\begin{tabular}{|r@{\extracolsep{0pt}.}l|r@{\extracolsep{0pt}.}l|r@{\extracolsep{0pt}.}l|}
\hline 
\multicolumn{2}{|c|}{Parameters} & \multicolumn{2}{c|}{Best Fit: New (old)} & \multicolumn{2}{c|}{Relative $\%$ Change in the Best Fit }\tabularnewline
\hline 
\multicolumn{2}{|c|}{$100~\omega_{b}$} & \multicolumn{2}{c|}{$2.236_{-0.038}^{+0.041}\left(2.24_{-0.04}^{+0.04}\right)$} & \multicolumn{2}{c|}{$0.2\%$}\tabularnewline
\multicolumn{2}{|c|}{$\omega_{{\rm cdm}}$} & \multicolumn{2}{c|}{$0.1176_{-0.0025}^{+0.0021}\left(0.117_{-0.002}^{+0.003}\right)$} & \multicolumn{2}{c|}{$0.5\%$}\tabularnewline
\multicolumn{2}{|c|}{$100\theta_{s}$} & \multicolumn{2}{c|}{$1.042_{-0.001}^{+0.001}\left(1.042_{-0.001}^{+0.001}\right)$} & \multicolumn{2}{c|}{$0.0\%$}\tabularnewline
\multicolumn{2}{|c|}{$\ln10^{10}A_{s}$} & \multicolumn{2}{c|}{$3.086_{-0.051}^{+0.046}\left(3.085_{-0.05}^{+0.047}\right)$} & \multicolumn{2}{c|}{$0.03\%$}\tabularnewline
\multicolumn{2}{|c|}{$n_{s}$} & \multicolumn{2}{c|}{$0.969_{-0.007}^{+0.010}\left(0.9726_{-0.0111}^{+0.0067}\right)$} & \multicolumn{2}{c|}{$0.4\%$}\tabularnewline
\multicolumn{2}{|c|}{$\tau_{{\rm reio}}$} & \multicolumn{2}{c|}{$0.07879_{-0.02672}^{+0.02461}\left(0.08009_{-0.02841}^{+0.02301}\right)$} & \multicolumn{2}{c|}{$1.6\%$}\tabularnewline
\multicolumn{2}{|c|}{$\Omega_{\Lambda}$} & \multicolumn{2}{c|}{$0.6989_{-0.0124}^{+0.0144}\left(0.7022_{-0.0157}^{+0.0111}\right)$} & \multicolumn{2}{c|}{$0.5\%$}\tabularnewline
\multicolumn{2}{|c|}{$Y_{{\rm He}}$} & \multicolumn{2}{c|}{$0.2478_{-0.0001}^{+0.0002}\left(0.2478_{-0.0001}^{+0.0002}\right)$} & \multicolumn{2}{c|}{$0.0\%$}\tabularnewline
\multicolumn{2}{|c|}{$H_{0}$} & \multicolumn{2}{c|}{$68.35_{-0.98}^{+1.13}\left(68.58_{-1.19}^{+0.92}\right)$} & \multicolumn{2}{c|}{$0.3\%$}\tabularnewline
\multicolumn{2}{|c|}{$\sigma_{8}$} & \multicolumn{2}{c|}{$0.8209_{-0.0198}^{+0.0174}\left(0.8194_{-0.0184}^{+0.0188}\right)$} & \multicolumn{2}{c|}{$0.2\%$}\tabularnewline
\multicolumn{2}{|c|}{$\Omega_{{\rm m}}$} & \multicolumn{2}{c|}{$0.301_{-0.014}^{+0.012}\left(0.2977_{-0.0111}^{+0.0158}\right)$} & \multicolumn{2}{c|}{$1.1\%$}\tabularnewline
\hline 
\end{tabular}
\end{table}

\begin{figure}[ht]
\includegraphics[width=14cm]{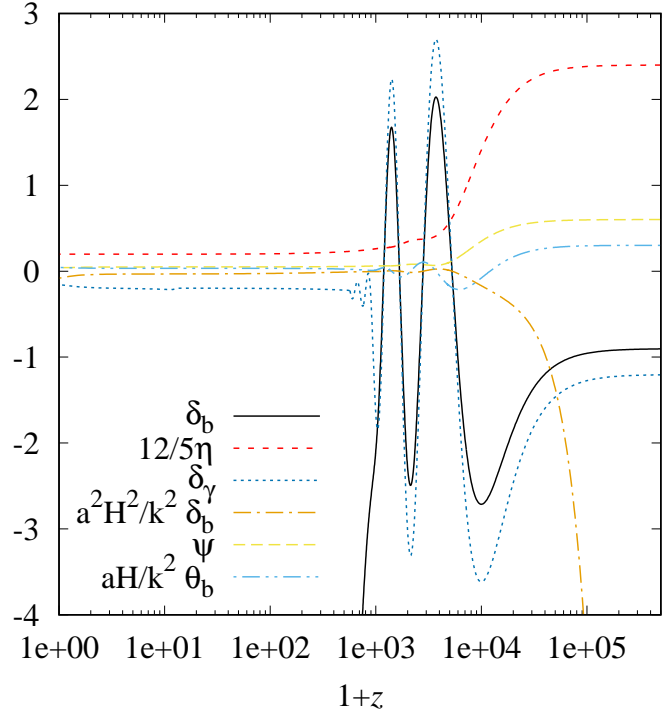} \caption{Evolution of perturbations out of which we can test the motivation
of our study in the high $k$ regime (fixing $k=0.1\ {\rm Mpc}^{-1}$)
at all times for the default values of the parameters. During radiation domination
(when $c_{s}^{2}$ should not be neglected) and up to $z>10^{4}$,
$\eta$ typically dominates over $\delta_{b}$, so that $c_{s}^{2}k^{2}\eta>c_{s}^{2}k^{2}\delta_{b}$,
and this term cannot be neglected. Instead, $\delta_{\gamma}$ dominates
over $\delta_{b}$ up to $z>10^{3}$, so that in this range of redshift
$c_{s}^{2}k^{2}\delta_{\gamma}>c_{s}^{2}k^{2}\delta_{b}$. Therefore, such
a term cannot be neglected in tight coupling approximation schemes.
Finally, even for this high $k$ mode, the subhorizon approximation
breaks down during radiation domination, as it is clear that for $z>5\times10^{4}$,
we have $c_{s}^{2}k^{2}\delta_{b}\ll c_{s}^{2}a^{2}H^{2}\delta_{b}$.
Since $c_{s}^{2}$ cannot be neglected in this redshift range, also
this term should be included in the equations of motion for the
perturbations. All these new terms are terms that are imposed by
conservation of the stress-energy tensor.}
\label{fig:perturbation_plot} 
\end{figure}

Now that we have the results, we are in the position to justify or
not the approximation taken on writing the non-covariant equations
of motion for the baryons, MB66--67. Once more, in those equations
and in all the ones that further make use of them, only the term $c_{s}^{2}k^{2}\delta_{b}$
was kept. As discussed before, this approximation leads to neglecting
several different terms, e.g.,\
$c_{s}^{2}k^{2}\delta_{\gamma},c_{s}^{2}k^{2}\eta,c^{2}k^{2}\psi,c_{s}^{2}a^{2}H^{2}\delta_{b}$.
This approximation is justified only if we can ignore all these terms
at all times at all scales. In particular, this implies that we should
have $\delta_{b}\gg\eta$ or $\delta_{b}\gg\delta_{\gamma}$ at all
times. To show that this approximation fails at times before dust
domination, we show in Figure\ \ref{fig:perturbation_plot} a plot
of the perturbations $\eta$, $\delta_{b}$, $\delta_{\gamma}$, $a^{2}H^{2}\delta_{b}/k^{2}$,
$\psi$ (this latter one being relevant for tight coupling approximation
in the Newtonian gauge), and $aH\theta_{b}/k^{2}$ in the high $k$ regime,
namely for $k=0.1\ {\rm Mpc}^{-1}$ at all times for the default values
of the parameters. We could see once and for all that keeping only the $c_{s}^{2}k^{2}\delta_{b}$
term was not a good approximation during radiation domination and up
to dust domination. Notice that this range of redshifts was exactly
the era in the universe at which we should not neglect $c_{s}^{2}$.

\begin{figure}[ht]
\includegraphics[width=9cm]{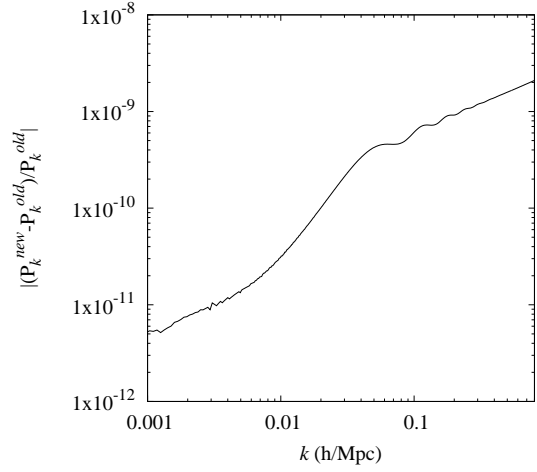} \caption{Relative difference in the matter power spectrum at $z=1000$. We fixed the same background parameters for both codes, so that
the error only depends on the difference in the equations of motion
for the baryon-photon sector.\label{fig:relative_mat_power_spec}}
\end{figure}

\begin{figure}[ht]
\includegraphics[width=14cm]{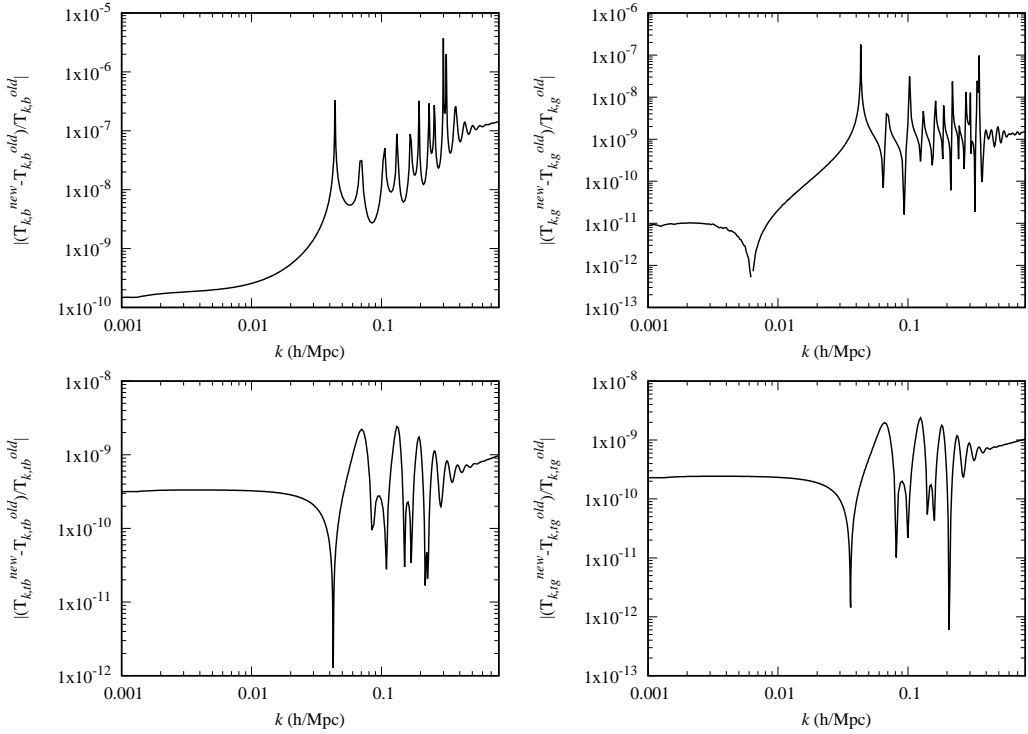} \caption{Relative difference in the transfer function for $\delta_{b}$, $\delta_{\gamma}$,
$\theta_{b}$, and $\theta_{\gamma}$ at $z=1000$. The $y$-axis gives
the relative error between covariant results and non-covariant results.
The error increases at higher values of $k$ and reaches values of
order $10^{-6}\sim10^{-5}$ for $\delta_{b}$.}
\label{fig:relative_transfer_function_z1000} 
\end{figure}

\begin{figure}[ht]
\subfloat[Relative difference in $\delta_{b}$]{\includegraphics[width=7cm]{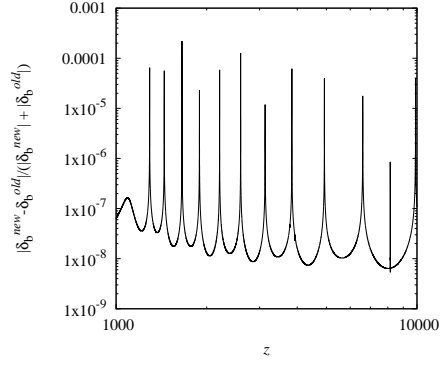}

}

\subfloat[Relative difference in $\delta_{\gamma}$]{\includegraphics[width=7cm]{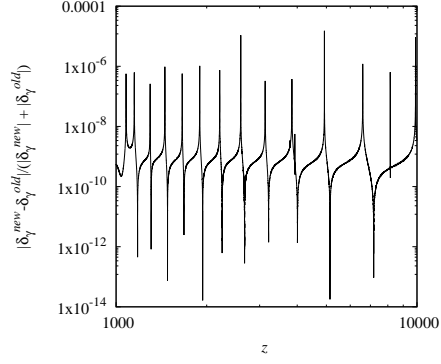}

}

\subfloat[Relative difference in $\eta$]{\includegraphics[width=7cm]{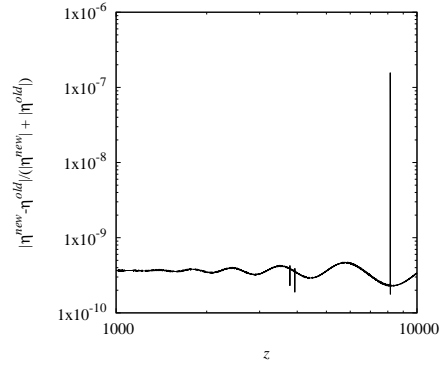}

}
\caption{Relative difference in various perturbation fields. The spikes are
due to oscillations crossing zero. In order to obtain these figures,
we increased the precision of CLASS (simply by setting a lower
tolerance for integrating the equations of motion). Furthermore, we
had to interpolate the numerical results in order to be able to evaluate
and compare the fields at the same point. \label{fig:rel_diff_pert}}
\end{figure}

Apart from the $1\%$ difference in the parameter estimation (which
in the end of the day, might be also correlated with finite-time parameter
sampling), we also studied the relative difference in the matter power
spectrum (see Figure\ \ref{fig:relative_mat_power_spec}) and various
transfer functions (see Figure\
\ref{fig:relative_transfer_function_z1000}), at high redshifts (we
chose $z=1000$), for which we expected the largest corrections (because
only at early times, we could consider the baryon to be non-pressureless).
Indeed, we could see a difference of up to $10^{-5}\sim10^{-6}$ for
the baryon energy-density transfer function between the two codes.
Since during early times $z\sim10^{4}$, second order tight coupling
approximation was of order $[aH/(n_{e}\sigma_{T}a)]^{2}\sim10^{-7}$,
then we could not neglect our corrections, as modern Boltzmann codes
arrived to implement also such an approximation scheme/precision. We
also give a relative difference for the perturbations $\delta_{b}$,
$\delta_{\gamma}$ and $\eta$ in Figure\ \ref{fig:rel_diff_pert}
. This ensured that the difference we saw was not merely numerical artifacts.
In other words, we think modern Boltzmann code should implement covariant
equations of motion in the baryon sector in order to be consistent
with second order tight coupling approximation schemes.

Furthermore, the results presented in this work and, e.g.,\ in Figure\
\ref{fig:perturbation_plot} were calculated only for $\Lambda$CDM.
If we were to study the phenomenology of some modified theory of gravity,
especially for early-time modified gravity theories (which in general
are implemented as not affecting directly the matter equations of
motion that come from conservation of the stress-energy tensor), we could
expect non-trivial behavior from the metric field perturbations $h,\eta$
(in the synchronous gauge) or $\phi,\psi$ (in the Newtonian gauge), which
could change the results of Figure\ \ref{fig:perturbation_plot}. A
priori, we would not be sure that the gravitational perturbation field
could be really neglected. This may result in missing relevant information
in studying the cosmology of modified theories of gravity. Keeping
equations that obey general covariance ensured we did not miss any
information within the theory.

\section{Conclusion\label{Conclusion}}

In this paper, we pointed out that there were at least three conceptual
issues in the implementation of the seminal work by Ma and Bertschinger
\citep{Ma:1995ey} in existing Boltzmann codes due to some missing 
terms. Those missing terms were numerically small and usually neglected,
but led to the following conceptual issues: (i) the baryon equations
were not gauge compatible; (ii) these equations violated the Bianchi
identity; (iii) the origin of the term with $c_{s}^{2}$ was not clear.
To address all these issues, we imposed the free Lagrangian of
baryons to be described by the covariant action of a non-relativistic
ideal gas. We found that this model for the baryon fluid, which described
a non-relativistic system of particles with a non-zero temperature
and $c_{s}^{2}$, led to covariant equations of motion for the perturbations,
which did not violate the Bianchi identities. We also derived
the same equations from the conservation of the stress-energy tensor,
without relying on the Lagrangian.

Since the new covariant equations of motion for baryons represent
one of the main results of our paper, we will rewrite them here both
in the synchronous and Newtonian gauge, in the notation of \citep{Ma:1995ey},
as follows.

{Synchronous gauge:}

\begin{subequations} 
\begin{align}
\dot{\theta}_{b} & =-\frac{\dot{a}}{a}\,\theta_{b}+c_{s}^{2}\,k^{2}\,\delta_{b}+R\,an_{e}\sigma_{T}(\theta_{\gamma}-\theta_{b})\\
 & \qquad{}+3\,c_{s}^{2}\,\frac{\dot{a}}{a}\,\theta_{b}\,,\label{eq:dottheta-1}
\end{align}
\label{eq:syncTH} \end{subequations} \begin{subequations} 
\begin{align}
\dot{\delta}_{b} & =-\theta_{b}-\frac{1}{2}\,\dot{h}\\
 & \qquad{}-\frac{6}{5}\,c_{s}^{2}\,\frac{\dot{a}}{a}\,\delta_{b}-\frac{3}{5}\,c_{s}^{2}\left(\theta_{b}+\frac{1}{2}\,\dot{h}\right).\;\;\;\label{eq:dotdelb-1}
\end{align}
\label{eq:syncDD} \end{subequations}

{Conformal Newtonian gauge:}

\begin{subequations} 
\begin{align}
\dot{\theta}_{b} & =-\frac{\dot{a}}{a}\,\theta_{b}+k^{2}\psi+c_{s}^{2}\,k^{2}\,\delta_{b}+R\,an_{e}\sigma_{T}(\theta_{\gamma}-\theta_{b})\\
 & \qquad{}+3\,c_{s}^{2}\,\frac{\dot{a}}{a}\,\theta_{b}\,,\label{eq:dottheta-2}
\end{align}
\end{subequations} \begin{subequations} 
\begin{align}
\dot{\delta}_{b} & =-\theta_{b}+3\dot{\phi}\\
 & \qquad{}-\frac{6}{5}\,c_{s}^{2}\,\frac{\dot{a}}{a}\,\delta_{b}-\frac{3}{5}\,c_{s}^{2}\,(\theta_{b}-3\dot{\phi}),\qquad\qquad\;\;\;\label{eq:dotdelb-2}
\end{align}
\end{subequations} where the new contributions (with respect to \citep{Ma:1995ey})
come from Lines (\ref{eq:dottheta-1}), (\ref{eq:dotdelb-1}),
(\ref{eq:dottheta-2}), and~(\ref{eq:dotdelb-2}).

Furthermore, with our new equations, we can keep terms that were before
completely omitted in the previous equations of motion. For example
on taking the time derivative of Equation\ (\ref{eq:syncDD}), which can
be written schematically as $\dot{\delta}_{b}=P$, we find $\ddot{\delta}_{b}+\mathcal{H}\dot{\delta}_{b}=\dot{P}+\mathcal{H}\,P$.
On replacing into this new equation the value of $\dot{\theta}_{b}$
with its own equation of motion, Equation~(\ref{eq:syncTH}), together
with the Einstein equations (to remove the fields $\ddot{h}$ and
$\dot{h}$), we get the standard term $k^{2}c_{s}^{2}\delta_{b}$,
but also other terms, e.g.,\ one proportional to $k^{2}\,c_{s}^{2}\,\eta$
($\eta$ being the curvature perturbation in the synchronous gauge), which
cannot be neglected a priori, but was absent in previous treatments
of baryon physics. In this same equation, also other terms have been
in the past neglected, e.g.,\ terms proportional to $a^{2}H^{2}c_{s}^{2}\delta_{b}$,
and from Equation\ (\ref{eq:dottheta-1}), a term proportional to $aHc_{s}^{2}\theta_{b}$.
Therefore, the approximations made in the past correspond to setting
$k\gg aH$, $\delta_{b}\gg\eta$, and $k^{2}\delta_{b}\gg aH\theta_{b}$.
However, during radiation domination (valid at least for the initial
conditions taken in \citep{Ma:1995ey} at $z\geq10^{6}$), on approximating
$H=H_{0}\sqrt{\Omega_{r0}}(1+z)^{2}$, where $H_{0}\approx h{\rm Mpc}^{-1}/(2997.9)$,
we find that $k$ is constrained to be $k\gg(1+z)\sqrt{\Omega_{r0}}/(3\times10^{3})\,h{\rm Mpc}^{-1}$.
Furthermore, $\eta\ll\delta_{b}$ (at any time and scale), and $k^{2}\delta_{b}\gg(1+z)\sqrt{\Omega_{r0}}/(3\times10^{3})\,h{\rm Mpc}^{-1}\,\theta_{b}$.
However, several authors, e.g.,\
\citep{Ma:1995ey}, considered a range for $k$ given by $0.01\,{\rm Mpc}^{-1}\leq k\leq10\,{\rm Mpc}^{-1}$,
so that inside this set, the previous approximations in general failed
for large redshifts. One may wonder then why our numerical results
were still similar to the previous ones. The reason was that $c_{s}^{2}$
was anyhow a small quantity, which gave only a small correction to
the baryon-dust fluid (i.e.,\ a fluid with $c_{s}^{2}$ identically
equal to zero). Nonetheless, we claim that only our approximation can
be trusted, as the $c_{s}^{2}$ contributions we introduced restored
general covariance and were the only terms consistent with cosmological
linear perturbation theory. We believe that the percent level correction
for the parameter estimation that we found is expected to be
important for future surveys \citep{Ade:2018sbj,Lazear_2014}.

With the covariant action so introduced, we claim we fixed all three issues of \citep{Ma:1995ey} stated above. In fact, the
new equations of motion for the baryon fluid possessed additional terms
of order $c_{s}^{2}$, which made the system of differential equations
gauge compatible, hence obeying the general covariance.

In order to understand the cosmological evolution at all relevant
redshifts, we need to study the solution of the equations of motion
before recombination. In this regime, photons and baryons are tightly
coupled, leading in general to a stiff regime during which it is difficult
to solve the equations of motion numerically. In order to overcome
this problem, we adapted several approximate schemes, already
introduced in the past, to our new dynamical equations of motion.
We then found the solutions of the equations of motion for the
perturbations up to the second order in the tight coupling approximation
using the new corrected baryon dynamics. In fact, we implemented
four different tight coupling approximation schemes without choosing
any gauge, so that our code could then be immediately used in any gauge
for modern Boltzmann solvers.

We therefore used a Monte Carlo sampler in order to re-estimate
the values of the cosmological parameters, after having incorporated
our covariant corrections into the baryon dynamical equations of motion.
We found that there were some parameters, e.g.,\ $\Omega_{{\rm m}}$,
whose best fit values deviated from the previous code analysis by one
percent. Moreover, we did not a priori know whether the deviation remained
as small as one percent in other models and various modified gravity
theories. In the age of precision cosmology, we therefore believe
that these changes are to be considered. On top of that, we found
that both the covariantly corrected baryon equations of motion and
the baryon-photon fluid tight coupling approximation schemes did not
make the code (implemented in CLASS) slower or stiffer than the previous
code. Hence, we did not find any reason why the modifications presented
in this paper to the code should not be included permanently into
modern Boltzmann codes, so as to confront any gravity theory with the
present and future observations in the era of precision cosmology.

\acknowledgments

We thank Julien Lesgourgues and Antony Lewis for suggestions and comments.
The work of S.~M.~was supported by Japan Society for the Promotion
of Science (JSPS) Grants-in-Aid for Scientific Research (KAKENHI)
No.\ 17H02890, No.\ 17H06359, and by World Premier International
Research Center Initiative (WPI), MEXT, Japan. M.~C.~P.~acknowledges
the support from the Japanese Government (MEXT) scholarship for Research
Student. The numerical computation in this work was carried out at
the Yukawa Institute Computer Facility.

\appendix

\section{Speed of Propagation of Matter Fields in the Presence of Several
Fluids\label{fluids}}

In the following, we want to consider the case of $N$ different fluids.
For simplicity, we will consider the case of several perfect fluids.
For each of the fluids, we need to give equations of state, namely: 
\begin{eqnarray}
p_{i} & = & p_{i}(\rho_{i},s_{i})\,,\\
T_{i} & = & T_{i}(\rho_{i},s_{i})\,,
\end{eqnarray}
which together with the first principle of thermodynamics, namely:
\begin{equation}
\text{d}\rho_{i}=\mu_{i}\text{d}n_{i}+n_{i}T_{i}\text{d}s_{i}\,,
\end{equation}
where $\mu_{i}\equiv(\rho_{i}+p_{i})/n_{i}$, are enough to completely
specify the thermodynamics of the $i^{\text{th}}$ fluid, provided the integrability
condition holds, namely: 
\begin{equation}
\left(\frac{\partial\mu_{i}}{\partial s_{i}}\right)_{n_{i}}=\left(\frac{\partial(n_{i}T_{i})}{\partial n_{i}}\right)_{s_{i}}.
\end{equation}
For any of the fluids, we have then automatically the speed of propagation, 
\begin{equation}
c_{s,i}^{2}\equiv\left(\frac{\partial p_{i}}{\partial\rho_{i}}\right)_{s_{i}}.
\end{equation}
So far, the discussion holds in {any} environment, in particular 
in cosmology. In the latter, the equations of state hold at any order
in perturbation theory, so that for relativistic degrees of freedom
$p=\rho/3$, not only at the level of the background, but at any order
in perturbation theory. In particular, at first order of perturbation
theory, we have: 
\begin{equation}
\delta p_{i}=\overline{\left(\frac{\partial p_{i}}{\partial\rho_{i}}\right)_{s_{i}}}\delta\rho_{i}+\overline{\left(\frac{\partial p_{i}}{\partial s_{i}}\right)_{\rho_{i}}}\delta s_{i}=\overline{c_{s,i}^{2}}\,\delta\rho_{i}+\overline{\left(\frac{\partial p_{i}}{\partial s_{i}}\right)_{\rho_{i}}}\delta s_{i}\,,
\end{equation}
where the overline tells us we need to consider the quantities evaluated
at the background level. Therefore, for fluids, whether or not in the
presence of other fluids, the expression of the speed of propagation
does not change. In particular, a dust fluid (for which $p=0=T$)
will have $c_{s,{\rm dust}}^{2}=0$, whereas for a radiation fluid
$c_{s,{\rm rad}}^{2}=1/3$. The thermodynamical description of the
fluid, well described in \cite{Misner:1974qy}, is powerful in particular
for non-barotropic fluids, such as ideal gas, for which $p$ is not
only a function of $\rho$. The speed of propagation in any fluid
does correspond to the adiabatic speed. It should be noticed that
this holds for fluids. In the case of a general action for a scalar
field, e.g.,\ quintessence, this formalism cannot be applied, and the
speed of propagation needs to be studied for the particular action
at hand.

We want to add here a discussion on the equations of state for baryons
in particular. One can choose several (but all equivalent to each
other) equations of state for the baryon fluid. In particular, one
can choose $p=p(T,\rho)$, which is absolutely fine, because a thermodynamical
equation of state can be written in terms of any two thermodynamical
degrees of freedom, for example $p=p(n,s)$ or $p=p(\rho,s)$ as discussed,
e.g.,\ on p.\ 564 of \cite{Misner:1974qy}. Then, it is obvious that
the pressure perturbation should be of the form: 
\begin{equation}
\delta p=\left(\frac{\partial p}{\partial\rho}\right)_{T}\delta\rho+\left(\frac{\partial p}{\partial T}\right)_{\text{\ensuremath{\rho}}}\delta T.
\end{equation}

Equivalently, we can consider as discussed in this paper $p=p\left(\rho,s\right)$.
These two descriptions are actually completely equivalent. In fact,
since the thermodynamical degrees of freedom are two, we can also
write $T=T(\rho,s)$, and hence, $\delta T=\left(\frac{\partial T}{\partial\rho}\right)_{s}\delta\rho+\left(\frac{\partial T}{\partial s}\right)_{\rho}\delta s$.
Then, one can show that: 
\begin{equation}
\delta p=\left(\frac{\partial p}{\partial\rho}\right)_{s}\delta\rho+\left(\frac{\partial p}{\partial s}\right)_{\rho}\delta s=\left[\left(\frac{\partial p}{\partial\rho}\right)_{T}+\left(\frac{\partial p}{\partial T}\right)_{\rho}\left(\frac{\partial T}{\partial\rho}\right)_{s}\right]\delta\rho+\left(\frac{\partial p}{\partial T}\right)_{\rho}\left(\frac{\partial T}{\partial s}\right)_{\text{\ensuremath{\rho}}}\delta s=c_{s}^{2}\,\delta\rho\,,
\end{equation}
for adiabaticity. This adiabaticity condition is also correctly considered
in Ma and Bertschinger, being stated before their Equation\
(96).

\section{Ideal Gas\label{IdealGas}}

One of the main points of our paper is to show indeed that the equations
of motion used in Boltzmann codes for the baryon fluid do not satisfy
energy-momentum conservation. Therefore, we find it appropriate to
enucleate the problem from another point of view. Each equation of
motion by itself defines automatically the system under study. Therefore,
one could try to understand what kind of matter is described by the
equations of motion for baryons used in Boltzmann codes.

To reach this goal, it is sufficient to consider these equations of
motion (MB66--67) and set to zero the interaction with the photon gas.
This should then define the self-gravitating physical system under
consideration. Then, the covariance breaking equations of motion in the
Newtonian gauge read as: 
\begin{align}
\dot{\delta}_{b} & =-\theta_{b}-3\dot{\zeta},\label{eq:MabaryondeltaFREE}\\
\dot{\theta}_{b} & =-\frac{\dot{a}}{a}\theta_{b}+k^{2}\alpha+c_{s}^{2}k^{2}\delta_{b}\,.\label{eq:MaBaryonthetaFREE}
\end{align}

The problems is that in this limit, the fluid does not reduce either to
a dust fluid or to any other perfect fluid equations of motions. The
free equations describe an unknown system, which cannot be described
in terms of a perfect fluid in general relativity. That is the reason
why these same equations break general covariance. In this same case,
we should also expect that these equations cannot come from a Lagrangian,
and therefore, they cannot be found by any conservation law, i.e.,\ they
cannot come from considering $T^{\mu\nu}{}_{;\nu}=0$.

Therefore, in order to reintroduce general covariance, we impose that
the free Lagrangian for the baryon reduces to the Lagrangian of an
ideal gas, because we are interested in a non-zero sound speed of
propagation, i.e.,\ $c_{s}^{2}\neq0$. In fact, a perfect fluid for
baryons may be modeled in two possible ways, with or without temperature.
The model without temperature is described by a dust fluid. However, this
model would lead to $c_{s}^{2}=0$, and this is not the model for which
we are looking. Let us consider then the case of a baryon fluid
with a tiny, but non-zero temperature, as only in this case, the baryon
fluid will possess $c_{s}^{2}\neq0$. The model for baryons considered
here consists of an ideal gas, whose fluid particles are considered
to be non-relativistic. To this fluid gas, a collision term with photons
will be then added, as we do in the case of a dust-like baryon-gas
without temperature.

A monoatomic ideal gas is defined by the following two equations of
state: 
\begin{equation}
p=n\,T\,,\qquad\rho=\mu_{g}\,n+\frac{3}{2}\,n\,T\,,\label{eq:eos}
\end{equation}
where $p,n,\rho$ and $T$ are the pressure, number density, energy
density, and temperature of the fluid, respectively. Furthermore, $\mu_{g}$
is a constant and represents the mass of the fluid particle. Since
the fluid is assumed to be non-relativistic, these equations of state
hold only for $T\ll\mu_{g}$. Indeed, as shown in Section 2.1, we
have performed an expansion of the equations of motion for the baryon
sector with respect to this small parameter.

The first law of thermodynamics for a general perfect fluid (see, e.g.,\ \cite{Misner:1974qy})
can be written as: 
\begin{equation}
\text{d}\rho=\mu\,\text{d}n+n\,T\,\text{d}s\,,
\end{equation}
where the enthalpy per particle $\mu$ is defined as: 
\begin{equation}
\mu=\frac{\rho+p}{n}\,,
\end{equation}
and $s$ represents the entropy per particle. Therefore, on considering
$\rho=\rho(n,s)$, we find: 
\begin{align}
\mu & =\left(\frac{\partial\rho}{\partial n}\right)_{s},\\
T & =\frac{1}{n}\left(\frac{\partial\rho}{\partial s}\right)_{n}.
\end{align}

On combining the two equations of state~(\ref{eq:eos}), it
is easy to show that: 
\begin{equation}
p=\frac{\rho T}{\mu_{g}+3T/2}\,,\label{eq:prhoT}
\end{equation}
which shows that an ideal gas does {not} represent a fluid with
a barotropic equation of state because $p\neq p(\rho)$; instead, we
have $p=p(\rho,T)$. Furthermore, this equation shows that for an
ideal gas $p/\rho=\mathcal{O}(T/\mu_{g})\ll1$, so that at the zeroth
order in $T/\mu_{g}$, this fluid can be well approximated by a dust
fluid. However, whenever in the history of the universe, on cosmological
scales, the speed of propagation for the baryon fluid cannot be neglected,
then we have\footnote{It should be noted that for a general perfect fluid, the notion of
the sound speed is purely thermodynamical, so that once the equations
of state are imposed, its expression is independent of the background.
For an ideal gas, we find $c_{s}^{2}=10T/(6\mu_{g}+15T)$. Furthermore,
since $p$ can be written in terms of two other thermodynamical variables,
we have that at linear order $\delta p=(\partial p/\partial\rho)_{s}\,\delta\rho+(\partial p/\partial s)_{\rho}\,\delta s$,
which can be rewritten as $\delta p=c_{s}^{2}\,\delta\rho+4\mu_{g}p/(6\mu_{g}+15T)\,\delta s$.
In particular, we find that, in general, $c_{s}^{2}\neq\delta p/\delta\rho$.}: 
\begin{equation}
c_{s}^{2}=\left(\frac{\partial p}{\partial\rho}\right)_{s}=\frac{\dot{p}}{\dot{\rho}}\approx\frac{\dot{\rho}T}{\mu_{g}\dot{\rho}}+\frac{\rho\dot{T}}{\mu_{g}\dot{\rho}}\approx\frac{T}{\mu_{g}}-\frac{a\dot{T}}{\mu_{g}(3\dot{a})}=\frac{T}{\mu_{g}}\left(1-\frac{1}{3}\,\frac{a}{T}\,\frac{\dot{T}}{\dot{a}}\right)=\frac{T}{\mu_{g}}\left(1-\frac{1}{3}\,\frac{d\ln T}{d\ln a}\right),\label{eq:speed_T}
\end{equation}
confirming Equation~(68) in \citep{Ma:1995ey} at the first approximation.
Hereafter, by an overdot, we represent the derivative with respect
to the conformal time $\tau$.

In particular, this shows that we cannot in general neglect the pressure
of such a fluid, and we have: 
\begin{equation}
p\approx\frac{T}{\mu_{g}}\,\rho\,.
\end{equation}

In the presence of a collision term with photons, there is an exchange
of entropy with the photon fluid, which when combined with the first
law of thermodynamics gives: 
\begin{equation}
\dot{T}=-2\,\frac{\dot{a}}{a}\,T+\frac{8}{3}\,\frac{\rho_{\gamma}}{\rho}\,\frac{\mu_{g}}{m_{e}}\,an_{e}\sigma_{T}(T_{\gamma}-T)\,,\label{eq:Tdot}
\end{equation}
as shown in \citep{Ma:1995ey}.

\section{Tight Coupling Approximation: Detailed Calculation\label{tcacal}}

Here, we discuss the tight coupling approximation up to the second
order with corrected equations of baryons.

We have the following set of equations for the photon fluid, without
fixing any gauge \cite{Ma:1995ey}\footnote{\label{footnote:3dcurvature}In the case of non-flat 3D slices, the
equations of motion need to be changed. For example, in Equation~(\ref{eq:s2_shear}),
the shear field gets an extra factor, $\sigma_{\gamma}\to s_{2}^{2}\,\sigma_{\gamma}$,
where, following the CLASS code notation, $s_{2}^{2}\equiv1-3K/k^{2}$.}: 
\begin{eqnarray}
\dot{\delta}_{\gamma} & = & -\frac{4}{3}\,\theta_{\gamma}-\frac{4}{3}\,\frac{k^{2}}{a}\,\chi+\frac{4}{3}\,\partial_{\tau}\!\left(\frac{k^{2}E}{a^{2}}\right)-4\dot{\zeta},\\
\dot{\theta}_{\gamma} & = & \frac{k^{2}}{4}\,\delta_{\gamma}-k^{2}\sigma_{\gamma}+k^{2}\alpha-\frac{1}{\tau_{c}}\,(\theta_{\gamma}-\theta_{b})\,,\label{eq:s2_shear}\\
2\dot{\sigma}_{\gamma} & = & \frac{8}{15}\left[\theta_{\gamma}+\frac{k^{2}}{a}\,\chi-k^{2}\,\partial_{\tau}\!\left(\frac{E}{a^{2}}\right)\right]-\frac{3}{5}\,k\,F_{\gamma3}-\frac{9}{5\tau_{c}}\,\sigma_{\gamma}+\frac{1}{10\tau_{c}}\,(G_{\gamma0}+G_{\gamma2})\,,\label{eq:dotsigmag}\\
\dot{F}_{\gamma l} & = & \frac{k}{2l+1}\left[lF_{\gamma\left(l-1\right)}-\left(l+1\right)F_{\gamma\left(l+1\right)}\right]-\frac{1}{\tau_{c}}F_{\gamma l}\,,\qquad l\ge3\\
\dot{G}_{\gamma l} & = & \frac{k}{2l+1}\left[lG_{\gamma\left(l-1\right)}-\left(l+1\right)G_{\gamma\left(l+1\right)}\right]\nonumber \\
 & & +\frac{1}{\tau_{c}}\left[-G_{\gamma l}+\frac{1}{2}\left(F_{\gamma2}+G_{\gamma0}+G_{\gamma2}\right)\left(\delta_{l0}+\frac{\delta_{l2}}{5}\right)\right]\,,\label{eq:dotG}
\end{eqnarray}
where $F_{\gamma2}=2\sigma_{\gamma}$, $F_{\gamma l}$ is higher multipoles
of the photon Boltzmann hierarchical equations, and $G_{\gamma l}$
is multipoles of Boltzmann hierarchical equations for the difference
in the photon linear polarization components \citep{Dodelson:2003ft,Ma:1995ey}.

We can rewrite the two equations for the speed of photons and baryons,
given respectively by (\ref{eq:s2_shear}) and (\ref{eq:dottheta}),
as: 
\begin{align}
\tau_{c}\left[\dot{\theta}_{\gamma}-\frac{k^{2}}{4}\,\delta_{\gamma}+k^{2}\sigma_{\gamma}-k^{2}\alpha\right]+\Theta_{\gamma b} & =0\,,\label{eq:eqthg1}\\
\tau_{c}\left[-\dot{\theta}_{b}-\frac{\dot{a}}{a}\,\theta_{b}+k^{2}\alpha+c_{s}^{2}\,k^{2}\left(\delta_{b}+3\,\frac{\dot{a}}{a}\,\frac{\theta_{b}}{k^{2}}\right)\right]+R\,\Theta_{\gamma b} & =0\,,\label{eq:eqthb1}
\end{align}
where\footnote{We have neglected the contribution from the baryon pressure from the
term $R=(\rho_{\gamma}+p_{\gamma})/(\rho_{b}+p_{b})$, because the
$c_{s}^{2}$ correction term, typically of order of $\Theta_{\gamma b}^{2}$,
will affect the tight coupling approximation only at higher orders
(e.g.,\ the cubic order).}: 
\begin{equation}
\Theta_{\gamma b}=\theta_{\gamma}-\theta_{b}\,.\label{eq:def_Theta_gb}
\end{equation}

Then, adding both of these equations, we obtain: 
\begin{equation}
\tau_{c}\left[\dot{\Theta}_{\gamma b}-\mathcal{H}\,\theta_{b}+c_{s}^{2}\,k^{2}\left(\delta_{b}+3\,\mathcal{H}\,\frac{\theta_{b}}{k^{2}}\right)-\frac{k^{2}}{4}\,\delta_{\gamma}+k^{2}\sigma_{\gamma}\right]+(1+R)\,\Theta_{\gamma b}=0\,,\label{eq:Theta_gammab}
\end{equation}
where $\mathcal{H}\equiv\frac{\dot{a}}{a}\,.$ The above equation
determines the evolution of $\Theta_{\gamma b}$, which is often referred
to as the slip parameter. Equation (\ref{eq:Theta_gammab}) involves
the shear of photons. The shear equation (\ref{eq:dotsigmag}) for
the photon can be rewritten as, 
\begin{equation}
\sigma_{\gamma}=\frac{\tau_{c}}{9}\left\{ \frac{8}{3}\left[\theta_{\gamma}+\frac{k^{2}}{a}\,\chi-k^{2}\,\partial_{\tau}\!\left(\frac{E}{a^{2}}\right)\right]-3\,k\,F_{\gamma3}-10\dot{\sigma}_{\gamma}\right\} +\frac{1}{18}\,(G_{\gamma0}+G_{\gamma2})\,.\label{eq:sigmaEqdot}
\end{equation}

We can consider linear combinations of Equations\ (\ref{eq:eqthg1}) and
(\ref{eq:eqthb1}) in order to eliminate $\Theta_{\gamma b}$, so
that we find: 
\begin{align}
\dot{\theta}_{b}+R\dot{\theta}_{\gamma} & =Rk^{2}\left(\frac{1}{4}\,\delta_{\gamma}-\sigma_{\gamma}\right)+(1+R)\,k^{2}\alpha-\mathcal{H}\left(1-3c_{s}^{2}\right)\,\theta_{b}+c_{s}^{2}\,k^{2}\delta_{b}\,.\label{eq:eqthb2}
\end{align}

From the above equation, we obtain the equation for $\dot{\theta}_{\gamma}$
as:
\begin{equation}
\dot{\theta}_{\gamma}=-\frac{\dot{\theta}_{b}}{R}+k^{2}\left(\frac{1}{4}\,\delta_{\gamma}-\sigma_{\gamma}\right)+\frac{1+R}{R}\,k^{2}\alpha-\frac{\mathcal{H}}{R}\left(1-3c_{s}^{2}\right)\,\theta_{b}+\frac{c_{s}^{2}\,k^{2}}{R}\delta_{b}\,.\label{eq:deltethg}
\end{equation}

Since $\dot{\Theta}_{\gamma b}=\dot{\theta}_{\gamma}-\dot{\theta}_{b}\,,$
we can rewrite Equation\ (\ref{eq:eqthb2}) as: 
\begin{equation}
\dot{\theta}_{b}+R(\dot{\Theta}_{\gamma b}+\dot{\theta}_{b})=Rk^{2}\left(\frac{1}{4}\,\delta_{\gamma}-\sigma_{\gamma}\right)+(1+R)\,k^{2}\alpha-\mathcal{H}\left(1-3c_{s}^{2}\right)\,\theta_{b}+c_{s}^{2}\,k^{2}\delta_{b}\,,
\end{equation}
or: 
\begin{align}
\dot{\theta}_{b} & =-\frac{1}{1+R}\left\{ \mathcal{H}\,(1-3c_{s}^{2})\,\theta_{b}-(1+R)\,k^{2}\alpha-Rk^{2}\left(\frac{1}{4}\,\delta_{\gamma}-\sigma_{\gamma}\right)-c_{s}^{2}\,k^{2}\delta_{b}+R\dot{\Theta}_{\gamma b}\right\} ,\label{eq:theta_b_0}\\
\dot{\theta}_{\gamma} & =\dot{\Theta}_{\gamma b}-\frac{1}{1+R}\left\{ \mathcal{H}\,(1-3c_{s}^{2})\,\theta_{b}-(1+R)\,k^{2}\alpha-Rk^{2}\left(\frac{1}{4}\,\delta_{\gamma}-\sigma_{\gamma}\right)-c_{s}^{2}\,k^{2}\delta_{b}+R\dot{\Theta}_{\gamma b}\right\} \nonumber \\
 & =-\frac{1}{1+R}\left\{ \mathcal{H}\,(1-3c_{s}^{2})\,\theta_{b}-(1+R)\,k^{2}\alpha-Rk^{2}\left(\frac{1}{4}\,\delta_{\gamma}-\sigma_{\gamma}\right)-c_{s}^{2}\,k^{2}\delta_{b}-\dot{\Theta}_{\gamma b}\right\} .
\end{align}

All these equations are exact. In what follows, we will mainly use
Equations\ (\ref{eq:def_Theta_gb}), (\ref{eq:Theta_gammab}), (\ref{eq:sigmaEqdot}),
and (\ref{eq:theta_b_0}) in order to find approximate solutions for
$\Theta_{\gamma b}$ and $\sigma_{\gamma}$.

\subsection{Terms in G
}

In the equation of motion for the shear, Equation\ (\ref{eq:dotG}), there
appear terms in $G_{\gamma0}$ and $G_{\gamma2}$. Let us first see
the perturbative solution of these two terms. Consider the equation
for $G_{\gamma l}$ with $l=1,3$, 
\begin{align}
\tau_{c}\dot{G}_{\gamma1} & =\frac{k\tau_{c}}{3}\,[G_{\gamma0}-2G_{\gamma2}]-G_{\gamma1}\,,\\
\tau_{c}\dot{G}_{\gamma3} & =\frac{k\tau_{c}}{7}\,[3G_{\gamma2}-4G_{\gamma4}]-G_{\gamma3}\,,
\end{align}
with the assumption, to be confirmed later on, that $G_{\gamma0}=\mathcal{O}(\tau_{c})$
and $G_{\gamma2}=\mathcal{O}(\tau_{c})$ and that $G_{\gamma4}$
is even more suppressed. In this case, let us look for solutions of
the kind: 
\begin{equation}
G_{\gamma1}=G_{\gamma1}^{(0)}+\tau_{c}G_{\gamma1}^{(1)}+\tau_{c}^{2}G_{\gamma1}^{(2)}\,.
\end{equation}

Then, we find: 
\begin{equation}
\tau_{c}[\dot{G}_{\gamma1}^{(0)}+\tau_{c}\dot{G}_{\gamma1}^{(1)}+\dot{\tau}_{c}G_{\gamma1}^{(1)}+\tau_{c}^{2}\dot{G}_{\gamma1}^{(2)}+2\tau_{c}\dot{\tau}_{c}G_{\gamma1}^{(2)}]=\mathcal{O}(\tau_{c}^{2})-(G_{\gamma1}^{(0)}+\tau_{c}G_{\gamma1}^{(1)}+\tau_{c}^{2}G_{\gamma1}^{(2)})\,,
\end{equation}
where we have assumed that $\dot{\tau_{c}}/(a\tau_{c})\simeq H$,
which is valid, as long as the tight coupling approximation is meaningful.
This last equation leads to the lowest order to: 
\begin{equation}
G_{\gamma1}^{(0)}=0\,.
\end{equation}

Then: 
\begin{equation}
\tau_{c}[\tau_{c}\dot{G}_{\gamma1}^{(1)}+\dot{\tau}_{c}G_{\gamma1}^{(1)}+\tau_{c}^{2}\dot{G}_{\gamma1}^{(2)}+2\tau_{c}\dot{\tau}_{c}G_{\gamma1}^{(2)}]=\mathcal{O}(\tau_{c}^{2})-(\tau_{c}G_{\gamma1}^{(1)}+\tau_{c}^{2}G_{\gamma1}^{(2)})\,,
\end{equation}
or, to the lowest order now: 
\begin{equation}
G_{\gamma1}^{(1)}=0\,,
\end{equation}
and: 
\[
\tau_{c}[\tau_{c}^{2}\dot{G}_{\gamma1}^{(2)}+2\tau_{c}\dot{\tau}_{c}G_{\gamma1}^{(2)}]=\mathcal{O}(\tau_{c}^{2})-\tau_{c}^{2}G_{\gamma1}^{(2)}\,,
\]
so that: 
\begin{equation}
G_{\gamma1}^{(2)}\neq0\,,
\end{equation}
and: 
\begin{equation}
G_{\gamma1}=\mathcal{O}(\tau_{c}^{2})\,.
\end{equation}

A similar argument leads to: 
\begin{equation}
G_{\gamma3}=\mathcal{O}(\tau_{c}^{2})\,.
\end{equation}

Now, we need to verify that indeed $G_{\gamma0}=\mathcal{O}(\tau_{c})$
and $G_{\gamma2}=\mathcal{O}(\tau_{c})$. In fact, we have: 
\begin{align}
\tau_{c}\dot{G}_{\gamma0} & =k\tau_{c}\,[-G_{\gamma1}]-G_{\gamma0}+\frac{1}{2}\,(2\sigma_{\gamma}+G_{\gamma0}+G_{\gamma2})\,,\nonumber \\
 & \approx-\frac{1}{2}\,G_{\gamma0}+\frac{1}{2}\,(2\sigma_{\gamma}+G_{\gamma2})\,,\\
\tau_{c}\dot{G}_{\gamma2} & =\frac{k\tau_{c}}{5}\,[2G_{\gamma1}-3G_{\gamma3}]-G_{\gamma2}+\frac{1}{10}\,(2\sigma_{\gamma}+G_{\gamma0}+G_{\gamma2})\,,\nonumber \\
 & \approx-\frac{9}{10}\,G_{\gamma2}+\frac{1}{10}\,(2\sigma_{\gamma}+G_{\gamma0})\,,
\end{align}
so that on looking for solutions of the kind $G_{\gamma0}=G_{\gamma0}^{(1)}+\tau_{c}G_{\gamma0}^{(2)}\,,$
and $G_{\gamma2}=G_{\gamma2}^{(1)}+\tau_{c}G_{\gamma2}^{(2)}\,,$
we find: 
\begin{align}
\tau_{c}\left[\dot{G}_{\gamma0}^{(1)}+\tau_{c}\left(\dot{G}_{\gamma0}^{(2)}+\frac{\dot{\tau}_{c}}{\tau_{c}}\,G_{\gamma0}^{(2)}\right)\right] & =-\frac{1}{2}\,(G_{\gamma0}^{(1)}+\tau_{c}G_{\gamma0}^{(2)})+\frac{1}{2}\,(2\sigma_{\gamma}+G_{\gamma2}^{(1)}+\tau_{c}G_{\gamma2}^{(2)})\,,\\
\tau_{c}\left[\dot{G}_{\gamma2}^{(1)}+\tau_{c}\left(\dot{G}_{\gamma2}^{(2)}+\frac{\dot{\tau}_{c}}{\tau_{c}}\,G_{\gamma2}^{(2)}\right)\right] & =-\frac{9}{10}\,(G_{\gamma2}^{(1)}+\tau_{c}G_{\gamma2}^{(2)})+\frac{1}{10}\,(2\sigma_{\gamma}+G_{\gamma0}^{(1)}+\tau_{c}G_{\gamma0}^{(2)})\,,
\end{align}
so that, at the lowest order, we find: 
\begin{align}
-\frac{1}{2}\,G_{\gamma0}^{(1)}+\frac{1}{2}\,(2\sigma_{\gamma}+G_{\gamma2}^{(1)}) & =0\,,\\
-\frac{9}{10}\,G_{\gamma2}^{(1)}+\frac{1}{10}\,(2\sigma_{\gamma}+G_{\gamma0}^{(1)}) & =0\,,
\end{align}
or: 
\begin{align}
G_{\gamma2}^{(1)} & =\frac{1}{2}\,\sigma_{\gamma}\,,\\
G_{\gamma0}^{(1)} & =\frac{5}{2}\,\sigma_{\gamma}\,.
\end{align}

Here, we assume for the moment that $\sigma_{\gamma}=\mathcal{O}(\tau_{c})$.
We will check later on that this assumption is consistent. Then, at
the next order: 
\begin{align}
\tau_{c}[\dot{G}_{\gamma0}^{(1)}] & =-\frac{1}{2}\,(\tau_{c}G_{\gamma0}^{(2)})+\frac{1}{2}\,(\tau_{c}G_{\gamma2}^{(2)})\,,\\
\tau_{c}[\dot{G}_{\gamma2}^{(1)}] & =-\frac{9}{10}\,(\tau_{c}G_{\gamma2}^{(2)})+\frac{1}{10}\,(\tau_{c}G_{\gamma0}^{(2)})\,,
\end{align}
or: 
\begin{align}
\frac{5}{2}\,\dot{\sigma}_{\gamma} & =-\frac{1}{2}\,G_{\gamma0}^{(2)}+\frac{1}{2}\,G_{\gamma2}^{(2)}\,,\\
\frac{1}{2}\,\dot{\sigma}_{\gamma} & =-\frac{9}{10}\,G_{\gamma2}^{(2)}+\frac{1}{10}\,G_{\gamma0}^{(2)}\,,
\end{align}
which leads to: 
\begin{align}
G_{\gamma0}^{(2)} & =-\frac{25}{4}\,\dot{\sigma}_{\gamma}\,,\\
G_{\gamma2}^{(2)} & =-\frac{5}{4}\,\dot{\sigma}_{\gamma}\,,
\end{align}
and to: 
\begin{align}
G_{\gamma0} & =\frac{5}{2}\,\sigma_{\gamma}-\frac{25}{4}\,\tau_{c}\dot{\sigma}_{\gamma}\,,\label{eq:solG0}\\
G_{\gamma2} & =\frac{1}{2}\,\sigma_{\gamma}-\frac{5}{4}\,\tau_{c}\dot{\sigma}_{\gamma}\,.\label{eq:solG2}
\end{align}

For general $l\geq3$, we have: 
\begin{equation}
\tau_{c}\dot{G}_{\gamma l}=\frac{k\tau_{c}}{2l+1}\,[lG_{\gamma,l-1}-(l+1)\,G_{\gamma,l+1}]-G_{\gamma l}\,,
\end{equation}
and since there are no source terms, we will assume that each $(l+1)^{\text{th}}$
term is suppressed by $\tau_{c}$ with respect to the $l^{\text{th}}$ term,
that is: 
\begin{equation}
G_{\gamma l}=\beta_{l}\,\tau_{c}G_{\gamma,l-1}\,,
\end{equation}
so that we find: 
\begin{equation}
\beta_{l}\,\tau_{c}^{2}\left[\dot{G}_{\gamma,l-1}+\frac{\dot{\tau}_{c}}{\tau_{c}}\,G_{\gamma,l-1}\right]=\frac{k\tau_{c}}{2l+1}\,[lG_{\gamma,l-1}-(l+1)\,\beta_{l+1}\,\tau_{c}G_{\gamma,l}]-\beta_{l}\,\tau_{c}G_{\gamma,l-1}\,,
\end{equation}
or: 
\begin{equation}
\beta_{l}\,\tau_{c}^{2}\left[\dot{G}_{\gamma,l-1}+\frac{\dot{\tau}_{c}}{\tau_{c}}\,G_{\gamma,l-1}\right]=\frac{k\tau_{c}}{2l+1}\,[lG_{\gamma,l-1}-(l+1)\,\beta_{l+1}\beta_{l}\,\tau_{c}^{2}\,G_{\gamma,l-1}]-\beta_{l}\,\tau_{c}G_{\gamma,l-1}\,,
\end{equation}
which leads, at leading order, to: 
\begin{equation}
0=\frac{k\tau_{c}}{2l+1}\,lG_{\gamma,l-1}-\beta_{l}\,\tau_{c}G_{\gamma,l-1}\,,
\end{equation}
or: 
\begin{equation}
\beta_{l}=\frac{kl}{2l+1}\,,
\end{equation}
so that we find 
\begin{equation}
G_{\gamma l}=\frac{l}{2l+1}\,k\tau_{c}\,G_{\gamma,l-1}\,,\qquad{\rm for}\;l\geq3.\label{eq:solGl}
\end{equation}

Since we obtain the same equations of motion for the terms $F_{\gamma l}$
for $l\geq3$, then we also have: 
\begin{equation}
F_{\gamma l}=\frac{l}{2l+1}\,k\tau_{c}\,F_{\gamma,l-1}\,,\qquad{\rm for}\;l\geq3.\label{eq:solFl}
\end{equation}

All Equations\ (\ref{eq:solG0}), (\ref{eq:solG2}), (\ref{eq:solGl}),
and (\ref{eq:solFl}) agree with the results given in \citep{Blas:2011rf}.

\subsection{Shear Solution}

Now, we need to look for an approximate solution for the shear. Using
the solutions for $G_{\gamma0,2}$ and $F_{\gamma3}$, we can rewrite
Equation~(\ref{eq:dotsigmag}) as: 
\begin{equation}
2\tau_{c}\dot{\sigma}_{\gamma}=\frac{8}{15}\,\tau_{c}\left[\theta_{\gamma}+\frac{k^{2}}{a}\,\chi-k^{2}\,\partial_{\tau}\!\left(\frac{E}{a^{2}}\right)\right]-\frac{3}{5}\,k\,\tau_{c}\,F_{\gamma3}-\frac{9}{5}\,\sigma_{\gamma}+\frac{1}{10}\left(\frac{5}{2}\,\sigma_{\gamma}-\frac{25}{4}\,\tau_{c}\dot{\sigma}_{\gamma}+\frac{1}{2}\,\sigma_{\gamma}-\frac{5}{4}\,\tau_{c}\dot{\sigma}_{\gamma}\right),
\end{equation}
which leads to: 
\begin{equation}
2\tau_{c}\dot{\sigma}_{\gamma}=\frac{8}{15}\,\tau_{c}\left[\theta_{\gamma}+\frac{k^{2}}{a}\,\chi-k^{2}\,\partial_{\tau}\!\left(\frac{E}{a^{2}}\right)\right]-\frac{3}{5}\,k\,\tau_{c}\,F_{\gamma3}-\frac{9}{5}\,\sigma_{\gamma}+\frac{1}{10}\left(3\,\sigma_{\gamma}-\frac{15}{2}\,\tau_{c}\dot{\sigma}_{\gamma}\right),
\end{equation}
and: 
\begin{align}
\left(2+\frac{3}{4}\right)\tau_{c}\dot{\sigma}_{\gamma} & =\frac{8}{15}\,\tau_{c}\left[\theta_{\gamma}+\frac{k^{2}}{a}\,\chi-k^{2}\,\partial_{\tau}\!\left(\frac{E}{a^{2}}\right)\right]-\frac{18}{35}\,k^{2}\,\tau_{c}^{2}\,\sigma_{\gamma}-\frac{3}{2}\,\sigma_{\gamma}\,,\nonumber \\
 & =\frac{8}{15}\,\tau_{c}\left[\theta_{\gamma}+\frac{k^{2}}{a}\,\chi-k^{2}\,\partial_{\tau}\!\left(\frac{E}{a^{2}}\right)\right]-\frac{3}{2}\,\sigma_{\gamma}+\mathcal{O}(\tau_{c}^{3})\,.
\end{align}

Now, let us assume that we have a solution of the form: 
\begin{equation}
\sigma_{\gamma}=\tau_{c}\sigma_{\gamma}^{(1)}+\tau_{c}^{2}\sigma_{\gamma}^{(2)}\,,
\end{equation}
then we obtain: 
\begin{equation}
\frac{11}{4}\,\tau_{c}^{2}\left[\dot{\sigma}_{\gamma}^{(1)}+\frac{\dot{\tau}_{c}}{\tau_{c}}\,\sigma_{\gamma}^{(1)}+\tau_{c}\left(\dot{\sigma}_{\gamma}^{(2)}+2\frac{\dot{\tau}_{c}}{\tau_{c}}\sigma_{\gamma}^{(2)}\right)\right]=\frac{8}{15}\,\tau_{c}\left[\theta_{\gamma}+\frac{k^{2}}{a}\,\chi-k^{2}\,\partial_{\tau}\!\left(\frac{E}{a^{2}}\right)\right]-\frac{3}{2}\,\left[\tau_{c}\sigma_{\gamma}^{(1)}+\tau_{c}^{2}\sigma_{\gamma}^{(2)}\right],
\end{equation}
which at the lowest order leads to: 
\begin{equation}
\sigma_{\gamma}^{(1)}=\frac{16}{45}\left[\theta_{\gamma}+\frac{k^{2}}{a}\,\chi-k^{2}\,\partial_{\tau}\!\left(\frac{E}{a^{2}}\right)\right],
\end{equation}
so that, at the next order, we have: 
\begin{equation}
\frac{11}{4}\,\tau_{c}^{2}\left[\dot{\sigma}_{\gamma}^{(1)}+\frac{\dot{\tau}_{c}}{\tau_{c}}\,\sigma_{\gamma}^{(1)}\right]=-\frac{3}{2}\,\tau_{c}^{2}\sigma_{\gamma}^{(2)}\,,
\end{equation}
or: 
\begin{equation}
\sigma_{\gamma}^{(2)}=-\frac{11}{6}\,\left[\dot{\sigma}_{\gamma}^{(1)}+\frac{\dot{\tau}_{c}}{\tau_{c}}\,\sigma_{\gamma}^{(1)}\right]=-\frac{88}{135}\,\frac{d}{d\tau}\left[\theta_{\gamma}+\frac{k^{2}}{a}\,\chi-k^{2}\,\partial_{\tau}\!\left(\frac{E}{a^{2}}\right)\right]-\left(\frac{11}{6}\,\frac{\dot{\tau}_{c}}{\tau_{c}}\right)\frac{16}{45}\left[\theta_{\gamma}+\frac{k^{2}}{a}\,\chi-k^{2}\,\partial_{\tau}\!\left(\frac{E}{a^{2}}\right)\right]\,.
\end{equation}

Hence, we find the approximate solution as: 
\begin{equation}
\sigma_{\gamma}=\frac{16}{45}\,\tau_{c}\left[\theta_{\gamma}+\frac{k^{2}}{a}\,\chi-k^{2}\,\partial_{\tau}\!\left(\frac{E}{a^{2}}\right)\right]\left(1-\frac{11}{6}\,\dot{\tau}_{c}\right)-\frac{88}{135}\,\tau_{c}^{2}\,\frac{d}{d\tau}\!\left[\theta_{\gamma}+\frac{k^{2}}{a}\,\chi-k^{2}\,\partial_{\tau}\!\left(\frac{E}{a^{2}}\right)\right].
\end{equation}

This approximate solution agrees with the one found in \citep{Blas:2011rf}.

\subsection{Slip Equation}

From now on, because we will make use of the new equations of motion
for the baryon fluid, our results will start differing from the ones
given in \citep{Blas:2011rf}. To find an approximate solution for
the slip parameter, up to the second order in $\tau_{c}$, let us
start with Equation\ (\ref{eq:Theta_gammab}): 
\begin{equation}
\tau_{c}\left[\dot{\Theta}_{\gamma b}-\mathcal{H}\,\theta_{b}+c_{s}^{2}\,k^{2}\left(\delta_{b}+3\,\mathcal{H}\,\frac{\theta_{b}}{k^{2}}\right)-\frac{k^{2}}{4}\,\delta_{\gamma}+k^{2}\left(\tau_{c}\sigma_{\gamma}^{(1)}+\tau_{c}^{2}\sigma_{\gamma}^{(2)}\right)\right]+(1+R)\,\Theta_{\gamma b}=0\,,
\end{equation}
and let us search for a solution for $\Theta_{\gamma b}$. Then, we
have, up to the second order: 
\begin{equation}
\tau_{c}\left[\dot{\Theta}_{\gamma b}^{(1)}+\dot{\Theta}_{\gamma b}^{(2)}-\mathcal{H}\,\theta_{b}+c_{s}^{2}\,k^{2}\left(\delta_{b}+3\,\mathcal{H}\,\frac{\theta_{b}}{k^{2}}\right)-\frac{k^{2}}{4}\,\delta_{\gamma}+k^{2}\left(\tau_{c}\sigma_{\gamma}^{(1)}+\tau_{c}^{2}\sigma_{\gamma}^{(2)}\right)\right]=-(1+R)\left[\Theta_{\gamma b}^{(1)}+\Theta_{\gamma b}^{(2)}\right].
\end{equation}

At the lowest order, we find: 
\begin{equation}
\tau_{c}\left[-\mathcal{H}\,\theta_{b}+c_{s}^{2}\,k^{2}\left(\delta_{b}+3\,\mathcal{H}\,\frac{\theta_{b}}{k^{2}}\right)-\frac{k^{2}}{4}\,\delta_{\gamma}\right]=-(1+R)\,\Theta_{\gamma b}^{(1)},
\end{equation}
or: 
\begin{equation}
\Theta_{\gamma b}^{(1)}=-\frac{\tau_{c}}{1+R}\left[-\mathcal{H}\,\theta_{b}+c_{s}^{2}\,k^{2}\left(\delta_{b}+3\,\mathcal{H}\,\frac{\theta_{b}}{k^{2}}\right)-\frac{k^{2}}{4}\,\delta_{\gamma}\right].\label{eq:sol_Theta_1}
\end{equation}

At the second order, we find: 
\begin{equation}
\tau_{c}\left[\dot{\Theta}_{\gamma b}^{(1)}+k^{2}\tau_{c}\sigma_{\gamma}^{(1)}\right]=-(1+R)\,\Theta_{\gamma b}^{(2)},
\end{equation}
or\footnote{If the background spatial curvature is present, then, as already mentioned
in Footnote \ref{footnote:3dcurvature}, we need to replace $\sigma_{\gamma}^{(1)}\to s_{2}^{2}\,\sigma_{\gamma}^{(1)}$.}: 
\begin{equation}
\Theta_{\gamma b}^{(2)}=-\frac{\tau_{c}}{1+R}\left[\dot{\Theta}_{\gamma b}^{(1)}+k^{2}\tau_{c}\sigma_{\gamma}^{(1)}\right].
\end{equation}

The solution is then found as: 
\begin{equation}
\dot{\Theta}_{\gamma b}=\dot{\Theta}_{\gamma b}^{(1)}+\dot{\Theta}_{\gamma b}^{(2)}\,.
\end{equation}

In the following, we will rewrite the above solution for the slip
parameter in such a way that we can easily implement them in the new
CLASS code.

\subsection{First Order Contribution}

We first manipulate the first order solutions as follows: 
\begin{align}
\dot{\Theta}_{\gamma b} & =\dot{\Theta}_{\gamma b}^{(1)}=\dot{\Theta}_{\gamma b}^{(1)}-\frac{\dot{\tau}_{c}}{\tau_{c}}\,\Theta_{\gamma b}^{(1)}+\frac{\dot{\tau}_{c}}{\tau_{c}}\,\Theta_{\gamma b}^{(1)}\,\nonumber \\
 & =\dot{\Theta}_{\gamma b}^{(1)}-\frac{\dot{\tau}_{c}}{\tau_{c}}\,\Theta_{\gamma b}^{(1)}+\frac{\dot{\tau}_{c}}{\tau_{c}}\,\Theta_{\gamma b}\nonumber \\
 & =\dot{\Theta}_{\gamma b}^{(1)}-\frac{\dot{\tau}_{c}}{\tau_{c}}\,\Theta_{\gamma b}^{(1)}+\frac{\beta_{1}\mathcal{H}}{1+R}\,\Theta_{\gamma b}^{(1)}-\frac{\beta_{1}\mathcal{H}}{1+R}\,\Theta_{\gamma b}^{(1)}+\frac{\dot{\tau}_{c}}{\tau_{c}}\,\Theta_{\gamma b}\nonumber \\
 & =\dot{\Theta}_{\gamma b}^{(1)}-\frac{\dot{\tau}_{c}}{\tau_{c}}\,\Theta_{\gamma b}^{(1)}+\frac{\beta_{1}\mathcal{H}}{1+R}\,\Theta_{\gamma b}^{(1)}-\frac{\beta_{1}\mathcal{H}}{1+R}\,\Theta_{\gamma b}+\frac{\dot{\tau}_{c}}{\tau_{c}}\,\Theta_{\gamma b}\,,\label{eq:slip_1_a}
\end{align}

In this equation, we replace the quantity $\dot{\theta}_{b}$ (which
appears inside $\dot{\Theta}_{\gamma b}^{(1)}$) with its solution
for Equation\ (\ref{eq:eqthb1}), $\dot{R}=-\mathcal{H}R$, $\dot{c}_{s}^{2}=\bar{c}_{s}^{2}-\mathcal{H}c_{s}^{2}$,
and the quantity $\Theta_{\gamma b}^{(1)}$ with Equation\ (\ref{eq:sol_Theta_1}).
It should be noticed that Equation\ (\ref{eq:eqthb1}) comes in the form:
\begin{equation}
\dot{\theta}_{b}=\dots+R\,\frac{\Theta_{\gamma b}}{\tau_{c}}\,.
\end{equation}

Therefore, when we replace $\dot{\theta}_{b}$, and we will end up with
an overall coefficient of $\Theta_{\gamma b}$ that partly depends
on $\dot{\tau}_{c}/\tau_{c}$ and partly on another coefficient that,
in turn, depends on the free function $\beta_{1}$. Then, we choose
$\beta_{1}$ so that the result looks as close as possible to the
result given in \citep{Blas:2011rf} (Equation~(2.19)), namely: 
\begin{equation}
\beta_{1}=2+R-3c_{s}^{2}R\,.
\end{equation}

\subsection{Second Order Contribution}

As we already know, at the second order, we find: 
\begin{equation}
\dot{\Theta}_{\gamma b}=\dot{\Theta}_{\gamma b}^{(1)}+\dot{\Theta}_{\gamma b}^{(2)}\,,\label{eq:D2_Theta_tot}
\end{equation}
where: 
\begin{align}
\Theta_{\gamma b}^{(1)} & =-\frac{\tau_{c}}{1+R}\left[-\mathcal{H}\,\theta_{b}+c_{s}^{2}\,k^{2}\left(\delta_{b}+3\,\mathcal{H}\,\frac{\theta_{b}}{k^{2}}\right)-\frac{k^{2}}{4}\,\delta_{\gamma}\right],\\
\Theta_{\gamma b}^{(2)} & =-\frac{\tau_{c}}{1+R}\left[\dot{\Theta}_{\gamma b}^{(1)}+k^{2}\tau_{c}\sigma_{\gamma}^{(1)}\right],\\
\sigma_{\gamma}^{(1)} & =\frac{16}{45}\left[\theta_{\gamma}+\frac{k^{2}}{a}\,\chi-k^{2}\,\partial_{\tau}\!\left(\frac{E}{a^{2}}\right)\right].
\end{align}

Therefore, we can substitute all the terms inside Equation\ (\ref{eq:D2_Theta_tot}).
This will lead to substituting $\ddot{\Theta}_{\gamma b}^{(1)}$,
$\dot{\Theta}_{\gamma b}^{(1)}$, $\dot{\sigma}_{\gamma}^{(1)}$,
$\sigma_{\gamma}^{(1)}$, $\ddot{R}=R\,(\mathcal{H}^{2}-\dot{\mathcal{H}})$,
$\dot{R}=-\mathcal{H}R$, ${c_{s}^{2}}'=\bar{c}_{s}^{2}-\mathcal{H}c_{s}^{2}$.

We then obtain a quite complicated expression that corresponds to
the needed answer for the slip parameter valid up to the second order
in $\tau_{c}$. However, once more, we try to write it as close as
possible to the expression written in \citep{Blas:2011rf}. Then, we
write: 
\begin{align}
\dot{\Theta}_{\gamma b} & =\dot{\Theta}_{\gamma b}^{(1)}+\dot{\Theta}_{\gamma b}^{(2)}-\left(1-\frac{2\mathcal{H}\tau_{c}}{R+1}\right)\frac{\dot{\tau}_{c}}{\tau_{c}}\,(\Theta_{\gamma b}^{(1)}+\Theta_{\gamma b}^{(2)})+\left(1-\frac{2\mathcal{H}\tau_{c}}{R+1}\right)\frac{\dot{\tau}_{c}}{\tau_{c}}\,(\Theta_{\gamma b}^{(1)}+\Theta_{\gamma b}^{(2)})\nonumber \\
 & =\dot{\Theta}_{\gamma b}^{(1)}+\dot{\Theta}_{\gamma b}^{(2)}-\left(1-\frac{2\mathcal{H}\tau_{c}}{R+1}\right)\frac{\dot{\tau}_{c}}{\tau_{c}}\,(\Theta_{\gamma b}^{(1)}+\Theta_{\gamma b}^{(2)})+\left(1-\frac{2\mathcal{H}\tau_{c}}{R+1}\right)\frac{\dot{\tau}_{c}}{\tau_{c}}\,\Theta_{\gamma b}\,.
\end{align}

In this last line, we should think that the quantities $\Theta_{\gamma b}^{(i)}$
are given explicitly, whereas $\Theta_{\gamma b}$ is left as it is.
Then, we still add new contributions: 
\begin{align}
\dot{\Theta}_{\gamma b} & =\left[\dot{\Theta}_{\gamma b}^{(1)}+\dot{\Theta}_{\gamma b}^{(2)}-\left(1-\frac{2\mathcal{H}\tau_{c}}{R+1}\right)\frac{\dot{\tau}_{c}}{\tau_{c}}\,(\Theta_{\gamma b}^{(1)}+\Theta_{\gamma b}^{(2)})\right]+\left(1-\frac{2\mathcal{H}\tau_{c}}{R+1}\right)\frac{\dot{\tau}_{c}}{\tau_{c}}\,\Theta_{\gamma b}\nonumber \\
 & +\frac{\text{\ensuremath{\beta_{2}}}}{R+1}\left(1-\frac{2\mathcal{H}\tau_{c}}{R+1}\right)\mathcal{H}\,(\Theta_{\gamma b}^{(1)}+\Theta_{\gamma b}^{(2)})-\frac{\text{\ensuremath{\beta_{2}}}}{R+1}\left(1-\frac{2\mathcal{H}\tau_{c}}{R+1}\right)\mathcal{H}\,(\Theta_{\gamma b}^{(1)}+\Theta_{\gamma b}^{(2)})\nonumber \\
 & +\beta_{3}\,\tau_{c}\,\Theta_{\gamma b}^{(1)}-\beta_{3}\,\tau_{c}\,\Theta_{\gamma b}^{(1)}\nonumber \\
 & =\left[\dot{\Theta}_{\gamma b}^{(1)}+\dot{\Theta}_{\gamma b}^{(2)}-\left(1-\frac{2\mathcal{H}\tau_{c}}{R+1}\right)\frac{\dot{\tau}_{c}}{\tau_{c}}\,(\Theta_{\gamma b}^{(1)}+\Theta_{\gamma b}^{(2)})\right]+\left(1-\frac{2\mathcal{H}\tau_{c}}{R+1}\right)\frac{\dot{\tau}_{c}}{\tau_{c}}\,\Theta_{\gamma b}\nonumber \\
 & -\frac{\text{\ensuremath{\beta_{2}}}}{R+1}\left(1-\frac{2\mathcal{H}\tau_{c}}{R+1}\right)\mathcal{H}\,(\Theta_{\gamma b}^{(1)}+\Theta_{\gamma b}^{(2)})+\frac{\text{\ensuremath{\beta_{2}}}}{R+1}\left(1-\frac{2\mathcal{H}\tau_{c}}{R+1}\right)\mathcal{H}\,\Theta_{\gamma b}\nonumber \\
 & +\beta_{3}\,\tau_{c}\,\Theta_{\gamma b}^{(1)}-\beta_{3}\,\tau_{c}\,\Theta_{\gamma b}\,,
\end{align}
where the functions $\beta_{2,3}$ are supposed to be of order $\mathcal{O}(\tau_{c}^{0})$,
and in the very last line, the equality holds up to the second order
in $\tau_{c}$. Therefore, we can rewrite: 
\begin{equation}
\dot{\Theta}_{\gamma b}=\mathcal{A}+\left(1-\frac{2\mathcal{H}\tau_{c}}{R+1}\right)\left(\frac{\dot{\tau}_{c}}{\tau_{c}}+\frac{\text{\ensuremath{\beta_{2}}}\mathcal{H}}{R+1}\right)\Theta_{\gamma b}-\beta_{3}\,\tau_{c}\,\Theta_{\gamma b}\,,
\end{equation}
where: 
\begin{equation}
\mathcal{A}\equiv\left[\dot{\Theta}_{\gamma b}^{(1)}+\dot{\Theta}_{\gamma b}^{(2)}-\left(1-\frac{2\mathcal{H}\tau_{c}}{R+1}\right)\frac{\dot{\tau}_{c}}{\tau_{c}}\,(\Theta_{\gamma b}^{(1)}+\Theta_{\gamma b}^{(2)})\right]-\frac{\text{\ensuremath{\beta_{2}}}}{R+1}\left(1-\frac{2\mathcal{H}\tau_{c}}{R+1}\right)\mathcal{H}\,(\Theta_{\gamma b}^{(1)}+\Theta_{\gamma b}^{(2)})+\beta_{3}\,\tau_{c}\,\Theta_{\gamma b}^{(1)}\,.
\end{equation}

Furthermore, we can easily find the decomposition of: 
\begin{equation}
\mathcal{A}=\mathcal{A}^{(1)}+\mathcal{A}^{(2)}\,,
\end{equation}
into the linear and quadratic contributions.

Let us then focus on $\mathcal{A}^{(2)}$. This term will contain
terms of the kind $\ddot{\theta}_{\gamma}$, $\dot{\theta}_{\gamma}$,
$\theta_{\gamma}$, and we replace them respectively by $\ddot{\theta}_{b}$,
$\dot{\theta}_{b}$, $\theta_{b}$ as their difference appears at
the cubic order. Once these terms are replaced, we can, in turn, replace
the expressions of $\dot{\theta}_{b}$ and $\ddot{\theta}_{b}$ by
using the zeroth order approximation found by using Equation\ (\ref{eq:theta_b_0}),
which can be written as: 
\begin{align}
\dot{\theta}_{b} & \approx\dot{\theta}_{b}^{(0)}\equiv-\frac{1}{1+R}\left[\mathcal{H}\,(1-3c_{s}^{2})\,\theta_{b}-(1+R)\,k^{2}\alpha-\frac{Rk^{2}}{4}\,\delta_{\gamma}-c_{s}^{2}\,k^{2}\delta_{b}\right],\\
\ddot{\theta}_{b} & \approx\ddot{\theta}_{b}^{(0)}\,.
\end{align}

This substitution is allowed because it is performed inside the quadratic
term $\mathcal{A}^{(2)}$. It can be checked that now in $\dot{\Theta}_{\gamma b}$,
there is no more explicit term containing $\theta_{\gamma}$ or any
of its derivatives. Furthermore, there is no more explicit dependence on $\ddot{\theta}_{b}$.
In the end, after all these substitutions, we have $\mathcal{A}^{(2)}\to\bar{\mathcal{A}}^{(2)}$.
However, in the linear contribution, we have a non-zero contribution
from the term $\dot{\theta}_{b}$. For such a term, we can replace
the exact solution coming from solving Equation\ (\ref{eq:eqthb1}), as
in: 
\begin{equation}
\dot{\theta}_{b}=-\mathcal{H}\,(1-3c_{s}^{2})\,\theta_{b}+k^{2}\alpha+c_{s}^{2}\,k^{2}\delta_{b}+\frac{R}{\tau_{c}}\,\Theta_{\gamma b}\,.
\end{equation}

This term will modify the coefficient of the term $\Theta_{\gamma b}$.
Then, we can choose the variables $\beta_{2,3}$ so that the linear
result looks as similar as possible to the one written in Equation~(2.20)
of \citep{Blas:2011rf}. Namely, we choose: 
\begin{align}
\beta_{2} & =-2-R+3Rc_{s}^{2}\,,\\
\beta_{3} & =\frac{2RH^{2}}{(R+1)^{2}}(1-3c_{s}^{2})\,,
\end{align}
so that in this case, the slip equation reduces to: 
\begin{equation}
\dot{\Theta}_{\gamma b}=\left(1-\frac{2\mathcal{H}\tau_{c}}{R+1}\right)\left(\frac{\dot{\tau}_{c}}{\tau_{c}}-\frac{2\mathcal{H}}{R+1}\right)\Theta_{\gamma b}+\mathcal{T}_{1}+\bar{\mathcal{A}}^{(2)}\,,
\end{equation}
so that the linear term (excluding the terms explicitly dependent
on $\Theta_{\gamma b}$) exactly coincides with the quantity $\mathcal{T}_{1}$.
We can finally add and subtract a quadratic order term $2\mathcal{H}\tau_{c}\mathcal{T}_{1}/(1+R)$
to end up with: 
\begin{equation}
\dot{\Theta}_{\gamma b}=\left(1-\frac{2\mathcal{H}\tau_{c}}{R+1}\right)\left[\left(\frac{\dot{\tau}_{c}}{\tau_{c}}-\frac{2\mathcal{H}}{R+1}\right)\Theta_{\gamma b}+\mathcal{T}_{1}\right]+\mathcal{T}_{2}\,,
\end{equation}
where: 
\begin{equation}
\mathcal{T}_{2}=\bar{\mathcal{A}}^{(2)}+\frac{2\mathcal{H}\tau_{c}}{R+1}\,\mathcal{T}_{1}\,.
\end{equation}

Here, the expressions for $\beta_{2,3}$ have been replaced in $\bar{\mathcal{A}}^{(2)}$.
Now, we are only left with the implementation of these results in
the CLASS code.

 \bibliographystyle{unsrt}
\bibliography{bibliography}

\end{document}